\documentclass[12pt,preprint]{aastex}

\newcommand{\sech}{\mbox{\rm sech}}

\newcommand{\kmskpc}{\,{\rm km}\,{\rm s}^{-1}\,{\rm kpc}^{-1}}

\newcommand{\pc}{\,{\rm pc}}
\newcommand{\kpc}{\,{\rm kpc}}

\newcommand{\mjy}{\,{\rm MJy}\,{\rm sr}^{-1}}

\shorttitle{Three Dimensional Structure of the Milky Way Disk}
\shortauthors{Drimmel \& Spergel}

\begin{document}

\title{Three Dimensional Structure of the Milky Way Disk: \\
the distribution of stars and dust beyond 0.35 $R_\sun$}

\author{Ronald Drimmel}
\affil{Osservatorio Astronomico do Torino,
    Pino Torinese, TO 10025, Italy}
\email{drimmel@to.astro.it}

\and

\author{David N. Spergel\footnote{Keck Distinguished Visiting
Professor of Astrophysics, IAS}}
\affil{Princeton University, Princeton, NJ  \& School of
Natural Sciences, IAS, Princeton, NJ}
\email{dns@astro.Princeton.edu}

\begin{abstract}

We present a three dimensional model for the Milky Way
fit to the
far-infrared (FIR) and near-infrared (NIR) data from the COBE/DIRBE
instrument for galactic latitudes $|b| < 30$ degrees and to within 20 degrees
of the Galactic center. Because of the low optical depth
at $240 \micron$, the FIR emission traces the distribution of Galactic dust
in the Galaxy.  We model the dust distribution as due to three components:
a warped exponential disk with scale length 0.28 $R_\sun$ and a flaring 
scale height, a spiral arm component
with four arms as traced by Galactic HII regions, and the local (Orion) 
arm which produces prominent emission features at galactic longitude
$l \simeq 80$ and $-100$ degrees. A Cosmic Infrared Background of $1.07\mjy$
is recovered, consistent with previous determinations. 
The dust distribution is then
used to calculate absorption in J and K, and the stellar
emission in these wavebands is modeled with two components: a warped
exponential disk with a scale length of 0.28 $R_\sun$ and a
spiral arm component dominated by {\em two} arms. This small
scale length is consistent with a maximal disk model for our Galaxy,
which is inconsistent with the cuspy dark matter halos predicted
in CDM models.
We find different amplitudes for the warp in the stars and dust, which
starts within the Solar Circle.


\end{abstract}


\section{Introduction}
%

The detailed study of the structure of the Milky Way enables astronomers
to address many of the most important questions in astrophysics, for it
is only in the Milky Way that we can make detailed studies
that enable us to infer the distribution of dark matter, the
star formation history of the Galaxy, the evolution of the Galaxy's
spiral structure and morphology.
The COBE satellite's all sky near- and  far-infrared maps
are a new window into the structure of our Galaxy and
the beginning of a bridge 
between optical and radio observations of the Milky Way.
Its potential in revealing the
Galaxy lies in the greater transparency of the interstellar medium (ISM)
to light at these longer wavelengths. The NIR emission arises mainly from
stars, as at visual wavelengths, but probes greater distances at low
galactic latitudes, while the FIR emission is an even more powerful probe
of Galactic structure as the Galaxy is nearly transparent at these 
wavelengths, much as in 
the radio regime. The FIR emission is from warm dust - the same dust that 
is responsible for absorption in the visual and NIR bands - and therefore
uniquely traces an important component of the ISM.

Due to extinction effects previous studies of Galactic structure
based on visual Galactic emission, and typically using star count
analysis, have been restricted to high galactic latitudes, and
therefore have been relatively insensitive to the non-axisymmetric 
structure in the stellar distribution. 
However, with NIR data large-scale 
asymmetry in the Galactic stellar distribution begins to be clearly
revealed in a Galactic warp and the barred structure of the Galactic bulge.
The existence of a Galactic warp has been known since the first
radio surveys of the Galaxy, but was not seen in the stellar
distribution until Djorgovski and Sosin's analysis of IRAS data \citep{DJO89}, 
and then later by \citet{FRE94} in the DIRBE NIR data.
Meanwhile the nonaxisymmetric structure of the Galactic bulge was first 
revealed in the IR by \citet{Blitz91}, and later confirmed in the 
NIR from COBE data by \citet{Weil94}.
Since these discoveries both parametric \citep{fred96,fre98}
and nonparametric methods \citep{GB96,BGS97} 
have been applied to COBE data to characterize these nonaxisymmetric features.

The Galactic bar and warp are now recognized features
of our Galaxy, yet another endemic nonaxisymmetric feature of
disk galaxies is spiral structure. In our Galaxy these are perhaps 
best mapped by HII regions \citep{GG76} and CO emission.
In this contribution we apply parametric models to both
FIR and NIR COBE data in order to describe the nonaxisymmetric
structure in our Galaxy beyond $r=0.35 R_\sun$. 

Other than the studies already mentioned above, 
the IR and FIR data has also been modeled (after the removal of stellar
emission in the IR by an assumed stellar emission model) 
by decomposing the emission
into two to three components, each associated with a different gas phase 
of the ISM, taken to be correlated to HI, CO, and HII radio fluxes
\citep[and references therein]{sod97}. 
Applying a dust emission model to the recovered spectra then
allows physical state variables of the dust to be inferred. This
approach efficiently allows the removal of the Galactic foreground emission 
to recover the Cosmic Infrared Background (CIB), and
sheds light on dust properties, but it does not yield a 
detailed map of the dust distribution; the decomposition procedure is
limited to describing gross radial variations of dust properties,
in part due to the kinematic distance ambiguity within the Solar Circle.

In contrast to the above approach we use a parametric model to 
characterize the spatial distribution of the dust, 
making as few assumptions as necessary regarding microscopic dust properties.
First efforts in this direction can be found in \citet{SMB97} and 
\citet{davies97}, who both compare the COBE/DIRBE FIR data with
the predicted emission from axisymmetric models of the dust distribution.
Here we employ a more complex model with
multiple flux density components including
not only an (cool) axisymmetric disk, but also (warm) nonaxisymmetric
features needed to account for major features in the FIR emission profile
of the Galaxy. We adjust the model parameters by fitting the
predicted $240 \micron$ sky emission to the $240 \micron$ Galactic
emission as seen in the COBE/DIRBE instrument.

Once a dust model is constructed using the FIR emission we
in turn use it to account for absorption in the 
NIR, and apply a parametric stellar flux density model to the 
DIRBE J and K band data. Like previous Galactic stellar distribution 
models, we employ an exponential disk component, but with the addition 
of spiral arms. A Galactic warp is applied to all components. Our efforts
are similar to that of \citet{fre98} (hereafter F98), though with 
some important 
differences, particularly in the derivation of the dust distribution
from FIR emission rather than from NIR absorption, as in \citet{SMB97}. 
We also limit our analysis to the Galactic plane ($|b| < 30$ deg), 
and exclude the Galactic center (GC) region ($|l| < 20\deg$), 
thereby obviating the need to model the Galactic bar and allowing us to focus
on the nonaxisymmetric structure at galactocentric radii $r > .35 R_\odot$.

In the following section the data reduction is described 
while in Section 3 the dust and stellar models are detailed. In
Section 4 the results of the parameter fitting procedure are given, 
and Section 5 then discusses the uncertainties of the 
parameters. The final two sections are 
reserved for discussion of our results in light of previous work,
and summarizes our conclusions.  

\section{Data}

In this work, we analyze the 'Zodi-Subtracted Mission Average (ZSMA)'
maps produced from the DIRBE data by \citet{dirbe2}
for the 240 $\mu$m band in the FIR, and the $J$ and $K$
bands in the NIR. 
The initial data set consists of a total of 393216 pixels 
in each waveband on a projected cube. Full skymaps of
this data are shown in 
\citet{dirbe3}. The data reduction procedures discussed below
were done using the {\small UIDL} data analysis package developed by
NASA's Goddard Space Flight Center, as well as additional 
{\small IDL} code developed to work in the {\small UIDL} environment.

From the NIR bands the point sources are first removed 
by applying a median filter to the data, without smoothing, to 
identify and reject point sources (pixels with much higher
intensities than neighbors) and questionable data points (pixels with much
lower intensities than neighbors). Using a 5 $\times$ 5 pixel window
a local median intensity is found at each pixel location, and pixels
are kept whose intensity is between 0.5 and 1.5 times the local median.
This procedure was applied to each NIR waveband separately, but any 
pixel rejected in one NIR band was rejected from the other NIR band as well.  
A local estimate of the uncertainty, or error, $\sigma_\lambda$, 
is made at each pixel location in each waveband ($\lambda = J,K$), 
via the standard deviation of the 
emission of non-rejected pixels in a 7 $\times$ 7 pixel window. 
The estimated error for the J band is shown in Figure \ref{jerror}
with respect to galactic latitude. The K band error map is similar.

Unlike the NIR emission, which comes primarily from unresolved point
sources, the FIR emission is inherently diffuse and possesses
structure at all angular scales. As a result the 
emission locally appears as diverse features such as ``ridges''
and ``peaks''. This necessitates that a different procedure be used to 
evaluate the local error. In addition, because the signal to noise is much
lower in the 240 $\mu$m band than in the NIR bands, pixels are found with 
negative intensities after Zodiacal light removal.
These values are obviously unphysical,
but we are want to simply exclude these data, as doing so would 
introduce a bias by effectively adding a net positive flux.
Some smoothing is required, but with care so as to not systematically 
redistribute flux near the Galactic plane (GP), where the 
gradient of the emission is high. 

At each pixel location
a second-order two-dimensional polynomial is fit to the FIR emission
in a 7 $\times$ 7 pixel region. The value of the resulting fit 
at the central pixel location is taken as the smoothed value,
while the standard deviation of the residuals 
in the window is used as an estimate of the local 
uncertainty, $\sigma_{240}$. This procedure introduces minimal smoothing 
with less noise than the data;
no negative intensities are present in the smoothed data, and the
total smoothed flux is equal to the total observed flux to 
two parts in ten thousand. It is to this smoothed data that our model of the 
the dust emission is fit. 
Figure \ref{fdat} shows a map of $\sigma_{240}$ and 
the ratio of the smoothed to raw data, demonstrating that 
no redistributed flux associated with the GP is present
in the smoothed data.  

Here some words concerning the purpose of a 
local estimation of the error are appropriate. 

The local errors are used in the calculation of the merit function 
used in the parameter fitting procedure to be discussed below (see Section 4). 
With COBE's large beam and its high sensitivities, we are in the ``confusion''
limit where stars just below our detection threshold and small scale
structures in the ISM are the dominant source of ``noise''.
By using local deviations from a local fit or local mean as a measure 
of the error, an empirical estimate is being made of how ``noisy'' the 
data is in each direction. This noise is due not only to instrumental
or measuring error, but also includes those variations arising from 
(often unresolved) structure on a scale smaller than the model intends to 
describe. In this sense ``uncertainty'' is a much better descriptor
than ``error''. For the NIR bands the primary source of uncertainty 
is unresolved stars, while in the FIR the uncertainty arises from 
both unresolved and resolved structure. In the merit function those 
pixels with higher error are given less weight than those with 
smaller error, and are thus allowed to deviate further from the model.

Finally, as we are primarily interested here in the distribution of stars 
and dust in the Galactic disk, all data with $|b| > 30^\circ$ are excluded.
Also excluded are pixels with longitudes within the 20 degrees of the 
GC, the Orion Nebula, and
regions centered on the extragalactic sources Andromeda, M33, and
the Magellanic Clouds. This leaves 
a total of 173569 pixels in the $240\micron$ dataset, while 
after point source removal there are 152371
pixels remaining in each NIR band, thus giving a total of 478311 data points
for the adjustment of the parameters.

\section{The Model}

\subsection{preliminaries}

Of the wavelengths at our disposal, the FIR bands 
provide the most direct indication of the distribution of Galactic dust,
whereas the NIR bands reflect the stellar distribution and give evidence 
of the dust distribution through both absorption and emission features. 
We first consider the nature of the FIR measurements before turning to 
the details of the model. 

The 240$\mu$m 
emission, after Zodiacal light has been subtracted, is due to 
Galactic dust and isotropic extragalactic emission. 
Using the same approach as \citet{SMB97}, we ignore
self-absorption or scattering at 240$\mu$m as the optical depth is
much less than one through the Galactic disk. The emission 
in any particular direction is then proportional to the 
dust column density along the line of sight and the dust's emissivity. 
Explicitly, the received specific intensity from a line-of-sight 
due to Galactic FIR emission is
\begin{equation}
\label{intensity}
I_\nu(T) = \int_0^\infty \rho(s) \epsilon_\nu(T) ~ds,
\end{equation}
where $\rho$ and $\epsilon_\nu$ represent the dust density and temperature
dependent specific emissivity respectively. 
However, the actual quantity represented in the
skymaps from the DIRBE instrument are inferred specific intensities based
on the assumption that $I_\nu\nu$ is constant. If the actual spectrum
is otherwise then a color correction term is needed to relate the 
inferred intensities to the actual intensities:
\begin{equation}
\label{colorcorr}
D_{\rm b} = I_{\rm b} K_{\rm b},
\end{equation}
where $K_{\rm b}$ is the correction term and $I_{\rm b}$ and $D_{\rm b}$ 
are the actual and inferred specific intensities at the representative 
frequency ${\rm b}$. The correction term 
\begin{equation}
\label{corrterm}
K_{\rm b}(T) = \int \frac{I_\nu}{I_{\rm b}} W_{\rm b}(\nu) ~d\nu ,
\end{equation}
$W_{\rm b}(\nu)$ being the normalized frequency response of the instrument 
for the b passband \citep{DIRBE}.

If the intensity arrives from a component with a single temperature, then
from equations (\ref{intensity}) and (\ref{colorcorr}) 
\begin{equation}
\label{onetcomp}
D_{\rm b} = \epsilon_{\rm b}(T) K_{\rm b}(T) ~\int \rho(s)ds 
    = \epsilon_{\rm b}(T) K_{\rm b}(T) n,
\end{equation}
with $n$ being the dust column density. If a sum of single temperature
components are present then
\begin{equation}
\label{sumcomp}
D_{\rm b} = \sum_i \epsilon_{\rm b}(T_i) K_{\rm b}(T_i) n_i = 
\sum_i k_{\rm b}(T_i) n_i,
\end{equation}
where 
\begin{equation}
\label{relem}
k_{\rm b}(T_i)=\frac{\epsilon_{\rm b}(T_i)K_{\rm b}(T_i)}
{\epsilon_{\rm b}(T_\circ)K_{\rm b}(T_\circ)}
\end{equation}
are relative emissivities, with the (unknown) constant 
$\epsilon_{\rm b}(T_\circ)K_{\rm b}(T_\circ)$ being absorbed into the 
column densities $n_i$. This constant remains unknown as long as
$T_\circ$ and $\epsilon_\nu(T)$ remain unspecified. Indeed, if
such a model for the emissive medium is assumed (as in the earlier 
work of \citet{SMB97}), it is sufficient
to define $k \equiv 1$ for one of the components to proceed, letting 
the absolute temperature and specific emissivity of the 
medium remain unknown.

In the case where the temperature is spatially varying a more 
complicated treatment is necessary. From equations (\ref{intensity}) through
(\ref{corrterm})
\begin{equation}
\label{dsource}
D_{\rm b} = \int I_\nu W_{\rm b}(\nu) ~d\nu 
= \int \rho(s) ~\int \epsilon_\nu(T) W_{\rm b}(\nu) ~d\nu ds
= \int \rho(s) \epsilon_{\rm b}(T) K_{\rm b}(T) ~ds,
\end{equation}
where we have defined a new (local) color correction term
\begin{equation}
\label{newcorr}
K_{\rm b}(T) 
= \int \frac{\epsilon_\nu(T)}{\epsilon_{\rm b}(T)} W_{\rm b}(\nu) ~d\nu.
\end{equation}
This new color correction term is spatially varying inasmuch as the
temperature varies spatially, but will be equivalent to the previous
color correction term (equation \ref{corrterm}) if the temperature is spatially
invariant. 

If it is assumed that the dust emissivity 
$\epsilon_\nu \propto \nu^\alpha B_\nu(T)$, where $B_\nu(T)$ is the Planck
function, then we can define a relative emissivity
\begin{equation}
\label{relemis}
k_{\rm b}(T) = \frac{B_{\rm b}(T)K_{\rm b}(T)}{B_{\rm b}(T_\circ)K_{\rm b}(T_\circ)},
\end{equation}
and write
\begin{equation}
\label{tvarsource}
D_{\rm b} = \int \rho(s) k_{\rm b}(T) ~ds,
\end{equation}
absorbing the constant $B_{\rm b}(T_\circ)K_{\rm b}(T_\circ)$ into the 
normalization of $\rho$.

It is important to point out that there is an inherent ambiguity
in scale when confronted solely with a 2D intensity skymap, 
which arises from the integrated flux column density of 3D emitting
structures. For arbitrary distributions of emitting matter, 
the relative size and flux density are mutually indeterminate; 
a given distribution will result in an intensity map 
identical to another with twice the size and half the flux density.
Therefore dimensionless units are used in the model by setting
$R_\odot \equiv 1$, effectively fixing the relative scale of the
disk and spiral arm components of our model.

In what follows $(r,\phi,z)$ represent galactic cylindrical
coordinates, $\phi$ being taken in the direction of Galactic rotation
and with the Sun lying along $\phi=0$ at $(R_\sun,0,Z_\sun)$.
The transformations used to
go from heliocentric galactic coordinates ($l,b,s$) to 
galactocentric Cartesian coordinates are 
\begin{equation}
\label{coords} 
\begin{array}{l} 
	x = R_\odot - s \cos(b) \cos(l) \\
	y = s \cos(b) \sin(l) \\
	z = s \sin(b) + Z_\odot. \\
	\end{array} 
\end{equation}

We are now prepared to describe the model of the dust distribution.

\subsection{dust emission model}

The inferred specific intensity from Galactic dust emission is modeled 
as arising from three density components, plus an isotropic contribution:
\begin{equation}
\label{idust}
D_{240}^{\rm mod}(l,b) = \int_0^\infty (k_{\rm d}\rho_{\rm axi} 
+ k_{\rm a}\rho_{\rm arm} + k_{+,-}\rho_{\rm loc}) ds + Q_{240} ~,
\end{equation}
where each component has an associated relative emissivity $k_j$, 
$j=d,a,+,-$. $Q_{240}$ is an offset applied to
account for isotropic extragalactic flux, i.e.
the cosmic infrared background (CIB).

The earlier work of \citet{SMB97} described the dust density with an 
axisymmetric exponential disk and implicitly assumed a constant emissivity 
(i.e. temperature); similarly, an exponential component is employed here
with the form 
\begin{equation}
\label{ddisk}
\rho_{\rm axi} = \rho_0 \exp (-r/h_r) \sech^2 (z/h_{\rm d}),
\end{equation}
where $h_r$ and $h_{\rm d}$ are the scale length and height respectively.
However, the scale height of the disk is given a linear flair, so that
\begin{equation}
\label{dflair}
h_{\rm d}(r) = \left\{ \begin{array}{ll} 
	h_{0,{\rm d}} +  h_{1,{\rm d}} (r - r_{\rm f})   & r > r_{\rm f} \\
	h_{0,{\rm d}} 		       & r \leq r_{\rm f} ~.
	\end{array} 
	\right.
\end{equation}
A hole is effected in the dust disk by taking 
$\rho_{\rm axi} \rightarrow \rho_{\rm axi}(r=0.5 R_\sun) 
\exp(-(r-0.5)^2/0.25^2)$ for $r < 0.5 R_\sun$. 
No inner dust ring is employed in the model.

For the disk component we assume a 
linear radial temperature gradient of $-6.8{\rm K}/R_\odot$ with a central
(reference) temperature of $T_\circ=26$K and minimum temperature 
of 3K. This temperature profile approximates that of the 
dust associated
with neutral hydrogen, as found by \citet{sod97}. 
From equation (\ref{relemis}) $k_{\rm d}(T)$ is computed,
using the fitted polynomial of 
\citet{Schleg98} for $K_{\rm b}(T)$,
consistent with the assumption that the dust emissivity is well 
described by $\epsilon_\nu \propto \nu^2 B_\nu(T)$
\citep{Draine84,dwek97}. For the range in temperatures assigned 
to the disk, the emissivity is approximately linear with temperature
out to 1.5$R_\odot$ (see Figure \ref{emprof}).

To this exponential disk component two additional single temperature
components are added, being needed 
to describe both prominent FIR emission and NIR absorption features,
namely the spiral arms of the Galaxy and a local feature, described as a 
spiral arm segment. Their respective emissivities are adjusted and 
relative to $k_{\rm d}(T_\circ)$.

For the spiral arms the geometry of \citet{GG76} is adopted, 
as implemented by \citet{TC93}, who map out four major spiral arms based on the
location of HII regions. The density profile across an arm is Gaussian, with 
a half-width $w_{\rm a} \propto r$ in the GP, and a flaring scale 
height $h_{\rm a}$. 
For any given position in the Galaxy the nearest point of the $i$th
spiral arm can be found numerically, with the corresponding minimum distance
$d_i, i=1,..,4$. (The enumeration of the arms follows that of Taylor and Cordes.)
The contribution of the spiral arm component to the density 
is taken to be from that arm for which the quantity $\exp-(d_i/w_{\rm a})^2$
is a maximum. We then have:
\begin{equation}
\label{arms}
\rho_{\rm arm} 
   = \rho_{\rm a} g_{\rm a} \exp-(d_i/w_{\rm a})^2 \exp -(z/h_{\rm a})^2 ~,
\end{equation}
where $\rho_{\rm a}$ is the density normalization for the arms and
$g_{\rm a}$ describes the radial cutoff of the density along the center
of the spiral arms, taken to be 
\begin{equation}
\label{garm}
g_{\rm a}(R_{\rm a}) = \left\{ \begin{array}{ll} 
	\exp(-(R_{\rm a} - r_{\rm m})^2/r_{\rm a}^2) & R_{\rm a} > r_{\rm m} \\
	1 		 	   & R_{\rm a} \leq r_{\rm m} ~,
	\end{array} 
	\right.
\end{equation}
$R_{\rm a}$ being the galactocentric radius of the nearest point of the spiral arm.
The scale width and height of the arms, $w_{\rm a}$ and $h_{\rm a}$, are 
functions of
$R_{\rm a}$; the width is assumed to be proportional to galactocentric radius
($w_{\rm a}=c_{\rm a} R_{\rm a}$) and the scale height of the arms is given a quadratic flair:
\begin{equation}
\label{hflair}
h_{\rm a}(R_{\rm a}) = \left\{ \begin{array}{ll} 
	h_{0,{\rm a}} + h_{1,{\rm a}} (R_{\rm a} - r_{\rm f,a})^2 
					& R_{\rm a} > r_{\rm f,a} \\
	h_{0,{\rm a}} 		 	& R_{\rm a} \leq r_{\rm f,a} ~.
	\end{array} 
	\right.
\end{equation}

To the description of the spiral arms we add a final refinement,
namely a reduction factor $f_{\rm r}$ on the size ($c_{\rm a}$ and $h_{\rm a}$)
of the $i=3$ (Sag--Car) spiral arm. 
The need for such a reduction factor can be seen immediately from the emission 
profile in the GP; the inequality of the emission peaks
at $\pm 50\deg$ longitude, arising from tangent points approximately
equidistant from the GC, show that the $i=2$ and 3 arms
are not equal.

The inner spiral arms account for emission features 
within $|l|\stackrel{<}{\sim}80\deg$ near the GP, producing peaks
in directions corresponding to the tangents of the arms, while the 
outer (Perseus) arm produces a broad emission feature from approximately 
$l=90$ to 180 degrees. 
However, inspection of the FIR emission profiles at low galactic latitudes
also shows two prominent features that lie at about $l=90$ and $-100$ 
degrees. These peaks 
correspond with prominent absorption features seen in the NIR data, 
showing them to be produced by local structure. The direction of
the features corresponds to the local (Orion) ``arm'' first seen 
in the distribution of young stars. While responsible for the 
majority of young stars in the vicinity of the Sun, the study of HII regions
by \citet{GG76} showed this local ``arm'' to be a minor feature relative 
to the major spiral arms in the Galaxy.

To model this local feature a spiral segment with a
gaussian density profile is employed: 
\begin{equation}
\label{local}
\rho_{\rm loc} = \rho_{\rm s} \exp(-d^2_{\rm s}/w^2_{\rm s})
\end{equation}
where $d^2_{\rm s} = (r - r_{\rm s})^2 + (z - Z_{\rm s})^2$, with 
$r_{\rm s} = R_{\rm s} \exp(-a_{\rm s} \phi)$ 
describing the spiral segment in the 
GP with a pitch angle $p_{\rm s}=\tan(a_{\rm s})$, 
and $Z_{\rm s}$ being the height of the local arm from 
the warped GP.
It was found necessary to place the Sun within a gap, 
parameterized by the azimuths of the gap boundaries, $\phi_1$ and $\phi_2$. 
That the Sun resides in a diffuse region
is also suggested by other studies of the local interstellar
medium \citep{pare84,fris95,fris96}. 
For computational convenience the local arm is truncated at the 
heliocentric distance where it becomes unresolved within a pixel. 
Beyond the truncation points and within the gap, $d_{\rm s}$ becomes the 
distance to the nearest end point of the local arm.
Finally, the local arm is given a different emissivity at positive verses negative 
longitudes ($k_{+,-}$), as it clearly shows evidence of having 
different temperatures: in emission it is much more prominent
at positive longitudes while its NIR absorption features at positive
and negative longitudes are very similar,
and the FIR color ratio ($D_{140}/D_{240}$) shows it to be 
significantly hotter at positive longitudes.

To resolve the ambiguity in scale the geometry of this local feature
must be fixed with respect to the GC. 
This is effected by fixing one of its geometrical parameters,
$\phi_1$, and adjusting the ratios 
$(w_{\rm s}/\phi_1)$, $(Z_{\rm s}/\phi_1)$, and $(\phi_2/\phi_1)$.
In principle
the additional information contained in the NIR absorption features 
associated with this local arm enables this structure to be accurately placed.

%
%
\subsection{stellar emission model}

We assume that the NIR flux is due to stellar emission, 
moderated by dust absorption:
\begin{equation}
\label{nirint}
\frac{dI_\nu}{ds} + \kappa_\nu \rho_{\rm d} I_\nu = \rho_* \epsilon_\nu,
\end{equation}
$\kappa_\nu$ representing the specific opacity of the dust
and $\epsilon_\nu$ the specific emissivity of the stars.
Integrating over frequency after multiplying by the passband's
frequency response function $W_{\rm b}(\nu)$, and remembering that
$D_{\rm b} = \int I_\nu W_{\rm b}(\nu) d\nu$, 
we arrive at the equation
\begin{equation}
\label{nirdb}
\frac{dD_{\rm b}}{ds} + \kappa_{\rm b} \rho_{\rm d} D_{\rm b} 
   = \rho_* \epsilon_{\rm b} K_{\rm b},
\end{equation}
where we have used the color correction factor $K_{\rm b}$ defined in
equation (\ref{newcorr}), though in this context it is not a function 
of temperature. The densities $\rho_{\rm d}$ and $\rho_*$ are for the dust and 
stars respectively, and the opacity is now a mean opacity:
\begin{equation}
\label{meanop}
\kappa_{\rm b}=\frac{\int \kappa_\nu I_\nu W_{\rm b}(\nu) ~d\nu}
                  {\int I_\nu W_{\rm b}(\nu) ~d\nu}.
\end{equation}
If the stellar NIR emission is dominated by a single stellar population,
possessing a characteristic specific emissivity $\epsilon_\nu$ (or 
luminosity function), then
$\epsilon_{\rm b}$ and $K_{\rm b}$ are spatially invariant. 
We therefore combine 
all the terms on the RHS of equation (\ref{nirdb}) into a stellar flux
density, $\eta_{\rm b}$, and arrive at
\begin{equation}
\label{istar}
D_{\rm b}^{\rm mod}(l,b)
=\int_{\rm los} \eta_{\rm b}(s) \exp(-\tau_{\rm b}(s)) ~ds + Q_{\rm b} ~.
\end{equation}
The mean optical depth $\tau_{\rm b}(s)$ is found from the 
dust distribution model and $Q_{\rm b}$ is an isotropic offset
term, albeit here without astrophysical justification.

The optical depth $\tau_{\rm b}(s)$ in equation (\ref{istar}) is defined as
\begin{equation}
\label{meantau}
\tau_{\rm b}(s) = \int_0^s \kappa_{\rm b} \rho_{\rm d} ~ds' ~.
\end{equation}
We then use the following approximation, 
treating $\kappa_{\rm b}$ as spatially invariant:
\begin{equation}
\label{tau}
\tau_{\rm b}(s) \approx \kappa_{\rm b} \int_0^s \rho_{\rm d} ~ds' 
= \kappa_{\rm V} (A_{\rm b}/A_{\rm V}) \int_0^s \rho_{\rm d} ~ds' ~.
\end{equation}
We adopt the ($A_{\rm J}/A_{\rm V}$) and ($A_{\rm K}/A_{\rm V}$) 
ratios of \citet{RL85}, 
and leave $\kappa_{\rm V}$ as an adjustable parameter. 
Rather than using the dust column density as given by the model
for calculating the opacity, we follow \citet{SMB97} and use 
a rescaled density, achieved here by applying a line-of-sight 
scaling factor to only one of the density components of the dust model. 
The scaling factor is determined by requiring that the predicted FIR emission
using the rescaled density be equal to the observed FIR emission. 
For a given line-of-sight the modeled FIR emission is
$D_{240}^{\rm mod} = \sum D_j + Q_{240}$, where the sum is over the
emission contributions of the three dust density components. The
appropriate scaling factor for component $j$ is then
\begin{equation}
\label{rescale}
f_j = \frac{D_{240}^{\rm obs} - \sum_{i \neq j}D_i - Q_{240}}{D_j} ~.
\end{equation}
For each line-of-sight one of the scale factors $f_j$ is chosen and 
used to rescale that component's density, i.e. 
$\widetilde{\rho_j} = f_j \rho_j$, and the total dust density 
$\rho_{\rm d}$ used in equation \ref{tau} is recomputed with this rescaled 
density. The component chosen to be rescaled is that whose scaling factor
results in the smallest fractional change in the component's 
column density, that is, the scale factor $f_j$ which minimizes $|1 - f_j|$.
However, if the modeled intensity from the nonaxisymmetric components
is at least 10\% of the observed intensity then either spiral arms or the 
local arm is rescaled. This bias only has affect in or near the GP where 
the condition is satisfied. 
It is also required that $f_j > 0$ to be considered as a valid rescaling
factor. This rescaling
procedure assumes that the entire residual in the FIR intensity,
$D_{240}^{\rm obs} - D_{240}^{\rm mod}$, is due to the modeled dust density 
deviating from the true density, rather than from deviations 
in the emissivity (i.e. temperature). 

If more than one stellar population is modeled, then the RHS of
equation (\ref{nirdb}) will consist of a sum of components, namely
$\sum_i \rho_{*,i} \epsilon_{{\rm b},i} K_{{\rm b},i} = \sum_i 
\eta_{{\rm b},i}$, where only the normalization of each component
is waveband dependent.  
The stellar flux density $\eta_{\rm b}$ is modeled using two components,
an exponential disk and a spiral arm
component, both with a $\sech^2$ vertical structure. 
The stellar spiral arms are assumed to have a density proportional
to the disk component, so that the total stellar flux
density is then 
\begin{equation}
\label{starem}
\eta_{\rm b} = \eta_{\rm b}^0 \exp(-r/r_*) \left[ \sech^2(z/h_*)
  + B_{\rm b} g(R_a) \exp-(d_i/w_{\rm a}^*)^2 \exp-(z/h_{\rm a}^*)^2 \right] ~.
\end{equation}
As for the dust, $d_i$ is the distance in the GP 
to the nearest spiral arm,
the half width $w_{\rm a}^*$ is proportional to $R_{\rm a}$, the radius of
the nearest point of the spiral curve, $h_{\rm a}^*$ is the scale 
height of the arms, assumed to be constant with galactocentric radius,
and $g(R_{\rm a})$ is a function that describes the variation in density
along the arm.
The parameter $B_{\rm b}$ (${\rm b = J,K}$) describes the relative 
amplitude of the arms;
allowing this parameter to vary with waveband is equivalent to assuming
that the arms consist of a stellar population that differs from that
of the disk component. In addition, a stellar cutoff factor is applied for
radii greater than the cutoff radius $r_{\rm c}$: 
\begin{equation}
\label{fcut}
f_{\rm cut} = \exp \left( -\frac{r-r_{\rm c}}{r_*/5.} \right) ~,
\end{equation}
applied so that $\eta_{\rm b} \rightarrow f_{\rm cut}\eta_{\rm b}$ 
for $r>r_{\rm c}$.

We explore two different basic geometries for the stellar spiral arms.
One is of a logarithmic form, whose phase and pitch angles are adjusted. 
In this case the $n^{\rm th}$ arm of $m$ logarithmic spirals has a radius
\begin{equation}
\label{lgspiral}
R_n = R_0\exp(-a\phi)\exp\left( a \frac{2\pi}{m} (n-1) \right) ~, n=1,..,m
\end{equation}
where $a$ is the tangent of the pitch angle, $p$.
For these spirals $g(R_{\rm a}) \equiv 1$.
Both $m=2$ and $m=4$ spiral geometries will be tried in
Section 5 against the data. 

The second spiral arm geometry, adopted for our standard model, 
is a sheared version of the spiral model employed
for the dust distribution. Under the assumption that
the dust spirals show the location of star formation fronts in the 
Galaxy, a young stellar population may be expected to drift ``downstream''
from the arm by an amount described by an offset in galactocentric azimuth. 
If the arms are assumed to have a fixed pattern in the presence of
a flat rotation curve, then the drift in azimuth is
\begin{equation}
\label{shear}
\phi_\tau = V_o \tau \left( \frac{1}{r} - \frac{1}{R_C} \right),
\end{equation}
where $V_o$ is the circular rotation speed, $\tau$ the mean age of the
population, and $R_C$ the corotation radius. The above shear in azimuth 
is applied to the loci of the points describing the dust spirals.
Because of mean color differences
in stellar populations of different ages, we expect the spiral arms
to have different offsets in the different wavebands. 
Thus we let $\tau$ vary with waveband (henceforth $\tau_{\rm b}$), as
well as the width of the arms (hence $w^*_{\rm b}=c^*_{\rm b}R_{\rm a}$).
In this formulation the geometry is found by adjusting parameters equivalent 
to $V_o \tau_{\rm b}$ and $R_C$. For these sheared spirals the function
$g(R_{\rm a})$ is the same as for the dust, and the reduction factor
on the Sag-Car arm is applied only to its width.
It is this model of the spiral arms that
we adopt as our standard model, shown in detail in the next section.

A global warp is added to all components by making the substitution
$z \rightarrow z'$ in the above formulation, where $z' = z - Z_{\rm w}$ 
for the dust components, or $z - Z_{\rm w}^*$ for the stellar components,
the function $Z_{\rm w}(r,\phi)$ describing the vertical displacement of
the warp. For the dust 
\begin{equation}
\label{zwarp}
Z_{\rm w} = h_{\rm w}(r) \sin(\phi - \phi_{\rm w}) ~,
\end{equation}
where $\phi_{\rm w}$ is the phase of the warp, and the amplitude function
\begin{equation}
\label{hwarp} 
h_{\rm w}(r) = \left\{ \begin{array}{ll} 
	a_{\rm w} (r - r_{\rm w})^2 	& r > r_{\rm w} \\
	0 		 	& r \leq r_{\rm w} ~,
	\end{array} 
	\right.
\end{equation}
with $r_{\rm w}$ being the galactocentric distance 
that the warp starts, and $a_{\rm w}$ an amplitude parameter. 
For the stars a separate amplitude coefficient $a_{\rm w}^*$ is 
applied, allowing the stellar warp to have a different
amplitude, similar to F98.

\section{Adjustment of parameters}

In order to determine the set of parameters that yields a best fit 
to the data we minimize a merit function of a Chi-squared form, 
specifically the sum of the $\chi^2$ for each waveband: 
\begin{equation}
\label{chi2}
\chi^2_{\rm b} = 
 \sum \left[ \frac{(D^{\rm mod}_{\rm b} - D^{\rm obs}_{\rm b})}{\sigma_{\rm b}} 
                \right]^2 ~,
\end{equation}
the second summation being over all pixels, and $\sigma_{\rm b}$ being
the uncertainties estimated in section 2. 
The general nonlinear minimization routine {\small N2FB} from 
the {\small PORT} mathematical library, 
developed by AT\&T Bell Laboratories, is used to minimize $\chi^2$.

The determination of the intensity $D^{\rm mod}_{\rm b}(l,b)$ predicted by a 
model entails the numerical evaluation of the line-of-sight
integrals in equations (\ref{idust}) for the FIR and (\ref{istar}) for the 
NIR. These numerical integrations are carried out by applying
the trapezoidal rule with an exponentially increasing
quadrature. The exponential convergence of the integrals allow us
to approximate the infinite range by integrating out to 
a finite number of effective scale lengths; we integrate to ten effective scale
lengths in one hundred steps. Intermediate values of the line-of-sight
integration through the dust are tabulated to determine the column density 
$\int \rho_{\rm d}(s) ~ds$ in equation (\ref{tau}) to calculate 
the optical depth $\tau(s)$.
To test the numerical integration, comparisons were made with
analytic solutions of an axisymmetric case, namely 
$\rho \propto \exp(-r/h_r - |z|/h_z)$, at various latitudes along the Galactic
meridian ($l=0,180$ degrees). The results of
these tests show a relative error of $2\times10^{-4}$.

Unfortunately due to computational limitations and the number of 
parameters that
must be employed to describe both the stars and the dust, a simultaneous
fit to all wavebands was not possible. 
The fitting of the parameters is thus done in a two step process. 
In the first step parameters describing
the dust distribution are adjusted using only the FIR data, 
then the remaining parameters to the NIR data.

\subsection{adjustment to the FIR}

The integration of the dust model along multiple lines-of-sight 
renders an FIR intensity skymap which can be directly compared
with the DIRBE $240 \micron$ skymap. However, there are 
several ambiguities which the FIR data alone does not allow to be
resolved. Already mentioned is the ambiguity in scale, which
leads us to adopt length units such that $R_\sun \equiv 1$.
A second ambiguity remains in the decomposition of the flux densities
into emissivities and densities, requiring that one or the other 
be fixed. For the fit to the FIR emission we specify that
$k_-=k_{\rm a}=1.$, which are relative to the
emissivity of the disk component, already defined by it's 
assumed temperature gradient. (The requirement that the local arm 
has the same density and dimensions at both positive and negative 
longitudes adds an additional constraint that allows $k_{+}$ to be 
adjusted relative to $k_{-}$.) 

While setting $R_\odot = 1$ defines the scale of the model,
the Sun's height above the GP, $Z_\odot$, in principle remains 
a free parameter to be adjusted. However, there is a near degeneracy 
between $Z_\odot$ and the other 
parameters that determine the latitude of the projected emission from
the local arm. For this reason it was found necessary to fix $Z_\odot$
in the FIR adjustment while adjusting the warp parameters and 
$(Z_{\rm s}/\phi_1)$. We have set $Z_\odot=0.001875$ ($15\pc/8\kpc$),
consistent with other estimates based on NIR data 
($16 \pc$, F98; $15.5 \pc$, \citet{Hamm95}), though smaller
than recent determinations based on optical starcount analysis 
($20.5 \pm 3.5 \pc$, \citet{Humph95}; $27 \pm 3 \pc$, \citet{Mendez98}).
However, \citet{HRCII} point out that these later estimates are sensitive
to the adopted scale height of the disk, and should be adjusted downward
to $\sim 15 \pc$ if a smaller effective scale height of the disk is 
adopted, as they advocate. It's worth noting that if the local arm is assumed 
to be in the GP (i.e. $Z_{\rm s}=0$), adjusting $Z_\odot$
leads to $Z_\odot < 0$, contrary to almost all studies of $Z_\odot$
which place the Sun above the GP; this was the primary
reason for introducing $Z_{\rm s}$ as a parameter.

A second parameter found to be poorly determined is the inner
scale height of the spiral arms, $h_{0,{\rm a}}$. This parameter is 
constrained by the emission of the inner spiral arm tangents,
but these make up too small a fraction of the total emission
to allow a reliable adjustment of $h_{0,{\rm a}}$. We have therefore set
$h_{0,{\rm a}} = 0.01 R_\odot$.

A total of twenty two parameters are adjusted in the FIR fit:
There are four adjusted parameters associated with the disk component
($\rho_0, h_{0,{\rm d}}, h_{1,{\rm d}}, h_r$), seven associated with 
the spiral arms ($\rho_{\rm a}, c_{\rm a}, h_{1,{\rm a}}, 
r_{\rm f,a}, r_{\rm m}, r_{\rm a}, f_{\rm r}$), seven associated with 
the local arm ($\rho_{\rm s}$, $R_{\rm s}$, $p_{\rm s}$, 
$(w_{\rm s}/\phi_1)$, $(Z_{\rm s}/\phi_1)$, 
$(\phi_2/\phi_1)$, $k_{{+}}$), three associated with the warp
($r_{\rm w}, a_{\rm w}, \phi_{\rm w}$) and the offset $Q_{240}$. 
Six parameters remain fixed: $\phi_1, r_{\rm f}, h_{0,{\rm a}}, Z_\sun$ and 
the emissivities $k_-$ and $k_{\rm a}$.
The resulting parameters of the preliminary dust fit are given in 
Table \ref{dustpars}, expressed for 
convenience in units where $R_\odot=8\kpc$ .

\placetable{dustpars}

We leave the general discussion of the implications of the parameters 
for later (Section 6), but here present the resulting skymap of the
FIR emission produced by the model as compared with the observations.
Figure \ref{fmaps} shows the observed and modeled $240\micron$ sky emission, 
and a map of their relative difference, 
$(D^{\rm obs}-D^{\rm mod})/D^{\rm obs}$.
Such grey scale maps give an overall impression of the emission,
but do not show the variations in the GP where most of the
details of interest are located. To show a more direct comparison between
the modeled and observed signal, emission profiles in the GP and
other galactic latitudes are shown in Figures \ref{firgp} and \ref{dprof}.
In these emission profiles it appears that the model has higher residuals
at higher latitudes. However, this is an effect of the
logarithmic scale used; Figure \ref{latdev} shows that the residuals are
in fact smaller at higher latitudes.

The emission profiles show that the major emission features are reproduced. 
In the GP the emission peaks at the
tangents of the spiral arms ($l \approx \pm 30, \pm 50$ and $-80\deg$), and
the peaks from the local arm (at $l \approx 80$ and $-100\deg$)
are correctly placed. In the skymaps the local arm shows itself
as two bright spots with considerable extent in latitude in nearly opposite 
directions of the sky. The Galactic warp is evident in the skymaps 
as an asymmetry in the Galactic emission at positive versus negative latitudes.
An emission feature due to the outer Perseus arm is evident as a broad feature
from approximately $l = 80$ to $-140\deg$ in Figure \ref{dprof}
for latitudes $b > 5\deg$. This arm is seen primarily at positive latitudes
due to the Galactic warp, and its large extent in latitude as
compared to the inner arms is what necessitates a flaring in the scale
height of the spiral arms.  The warp also accounts for the
small negative deviation in latitude at $l \approx -80 \deg$, 
a feature associated with dust within the Solar Circle.

\subsection{adjustment to the NIR}

A final adjustment to the NIR data yields the remaining parameters of 
the stellar distribution and resolves the remaining uncertainties 
of the dust distribution; the additional information provided by
NIR absorption, proportional to the dust column density,
allows the decomposition of the dust densities and FIR emissivities. 
For example, fixing all other parameters of the spiral arms, the FIR 
flux column density from the arms arriving from any particular 
line-of-sight is simply proportional to the product ($k_{\rm a}\rho_{\rm a}$). 
The decomposition is effected by adjusting $\rho_{\rm a}$
and varying the emissivity $k_{\rm a}$ so as to keep the product 
($k_{\rm a}\rho_{\rm a}$) constant, thus preserving the modeled FIR signal. 
Similarly for the local arm, the column flux density is fixed while
$\rho_{\rm s}$ is adjusted. 

In the fit to the NIR the parameter $Z_\odot$ is adjusted, while the distance 
to the warped plane, $Z_\odot - Z_{\rm w}({\bf x}_\odot)$, is preserved 
by appropriately varying $\phi_{\rm w}$, leaving the parameters 
$a_{\rm w}$ and $R_{\rm w}$ fixed to the values found in the FIR fit.
Meanwhile, the amplitude factor for the stars, $a_{\rm w}^*$, is adjusted to
allow the stellar warp to have a different amplitude than that found
in the dust. 

In summary there are nineteen adjusted parameters in the adjustment
to the NIR data:
five parameters describing with the disk component
($\eta^0_{\rm J,K},r_*, h_*, r_{\rm c}$), 
seven parameters associated with the spiral arms
($B_{\rm J,K}, c^*_{\rm J,K}, V_o\tau_{\rm J,K},R_{\rm C}$),
three parameters associated with the dust and it's absorption 
($\kappa_{\rm V},\rho_{\rm a},\rho_{\rm s}$), 
the stellar warp amplitude parameter, $a_{\rm w}^*$,
the vertical coordinate of the Sun, $Z_\odot$,
and the offsets $Q_{\rm J,K}$.
The stellar model possesses only a single parameter that is explicitly fixed,
the scale height of the stellar spiral arms, 
set to be equivalent to that of the dust ($h^*_{\rm a} = h_{0,\rm a}$).
However, several assumptions are made with regards to common geometry
between the dust and the stellar distributions, such as the radius that the
warp starts. 

The parameters resulting from the NIR adjustment are given in Table 
\ref{starpars}. 
The NIR skymaps of the DIRBE instrument and of the model are given in Figures
\ref{threej} and \ref{threek}, as well as the relative differences between 
the data and the model, $(D^{\rm obs}-D^{\rm mod})/D^{\rm obs}$. 
The skymaps of the observations and model
look very similar, the deviations only being obvious in the relative
difference maps. These show that the largest deviations are in the GP
within $30 \deg$ of the GC. Again, to show in more 
detail the concordance and deviations between the observed and the modeled
emission, especially in and near the GP, emission profiles at various 
latitudes are shown in Figures \ref{nirgp}, \ref{jprof} and \ref{kprof}.

\placetable{starpars}

The observed NIR emission profiles are much smoother than those seen in 
the FIR, and in contrast to the FIR the axisymmetric component dominates the 
emission profiles at all latitudes. The profiles deviate most from an ideal 
axisymmetric
profile as one approaches the GP, especially in J, and this is largely due 
to absorption. The spiral arms are much less evident
than in the FIR, and only provisionally identifiable in the K band
\citep{Dri00}. Indeed, differentiating between emission and absorption
features in the NIR is problematic, and it has been suggested that 
deviations from axisymmetry in the NIR can be attributed to absorption effects
alone \citep{Kent91}. 
The two absorption features most clearly evident are due to 
the local arm, seen at $l \simeq 80$ and $-100 \deg$.

In the modeled emission profiles the effect of using a rescaled dust
model to calculate extinction is immediately obvious, 
introducing fluctuations even on small scales. Incidences can be found
where the rescaling introduces spurious features, but more
often it allows the model to achieve a correspondence with the data
that could not be obtained otherwise. This 
is particularly evident at low latitudes in J, and shows the importance
of absorption at these latitudes in shaping the profiles. 
Nevertheless, emission from the spiral arms is important 
in the GP, though we will have to make comparisons 
with axisymmetric models in Section 5 to make this demonstrably obvious.

One nonaxisymmetric feature that is evident in the NIR is the Galactic warp,
though less obvious than in the FIR. 
It reveals itself as an asymmetry in the
emission profiles, seen most clearly in Figure \ref{kprof} 
at $|b|=5\deg$; the profiles are skewed toward positive longitudes at
$b=-5\deg$, and toward negative longitudes at $b=5\deg$. 
This requires a local tilt in the stellar distribution with respect to
the $b=0$ plane, accounted for here by the warp starting within 
the Solar Circle.

The relative difference map in K suggests the presence of
structure not accounted for by the model. 
Aside from the deviations in the GP mentioned 
above, broad bright patches in the difference map in K can be seen at 
positive and negative longitudes. That at positive longitudes, roughly
located at $l=90\deg$, is distinctly above the GP, and may be stellar emission
associated with the local (Orion) arm. The bright patch at negative
longitudes is less obvious, centered on the GP and 
is found within $l=-90\deg$, thus
not correctly positioned to be associated with the local arm.
Additional features seen in the difference map in J are associated with 
over or under estimated absorbtion.

\subsection{General features of the model}

To concisely summarize the general features of the standard model a
``bird's-eye-view'' of the Milky Way is presented as a surface density map
for the dust in Figure \ref{dmap}, and as a K band surface brightness map in 
Figure \ref{smap}. The center of the maps are determined from an 
extrapolation of
the model, which is itself specifically constructed only for $r > .34 R_\sun$,
thus the picture presented for $r < .34 R_\sun$ is necessarily incomplete. 
In particular, no structure associated with the Galactic bulge
is represented. Also, the spiral arms are incomplete on the far-side
of the Galaxy due to the adopted spiral geometry based on the 
{\em observed} HII regions. The relative
strength of the spiral arms in the dust and the stars is shown in Figure 
\ref{intarm} as the arm--interarm (flux) density contrast in the GP.
However, the difference in the scale heights of the disk and spiral arms 
mitigate the surface brightness/density contrast between the arm and 
interarm regions; the arm--interarm ratio in the NIR
surface brightness is 1.2 and 1.32 for J and K respectively.
Figure \ref{scaleh} shows the scale heights of the various components. 

In addition to the major spiral arms
the dust surface density map 
shows the smaller local arm in the region of the Sun.
Though the local arm produces prominent FIR emission
features in our sky, from an extra-Galactic perspective it is revealed to 
be a minor feature, as pointed out by \citet{GG76}. 
Indeed, ``arm'' is perhaps a misnomer. Similar 
structures are undoubtedly found throughout 
the Galaxy in the form of spurs and bridges between the main spiral arms; 
the local arm is prominent and resolved in our FIR sky only 
by virtue of its vicinity. 

Our flux density model implies an extinction corrected
K magnitude of $-$23.79 for the Milky Way, assuming $R_\sun = 8. \kpc$,
of which the spiral arm component contributes 7 percent of the total
luminosity. However, our spiral arm model is incomplete for the side
of the Galaxy opposite the Sun due to the lack of observed HII regions. 
Assuming that the spiral arm model is complete over 3/4 of the 
Galaxy, the actual contribution of the spiral arms to the 
total flux would be 10 percent, resulting in a K magnitude of $-$23.82, which
does not include light from the Bulge. The bulge/disk luminosity ratio 
has been approximately estimated as 0.2 \citep{Ortpriv}, which results 
in a final K magnitude of $-$24.02 for the Milky Way,
consistent with the earlier estimate of \citet{Mal96}.

\section{Uncertainties and alternative models}

In this section we explore the sensitivity of the adjusted
parameters on the choice of fixed parameters, thereby estimating 
systematic errors, and test the relative importance of specific 
features of the model that are not obviously essential. In particular
various spiral models for the stars are considered in an
attempt to reproduce the observed NIR emission profiles in the GP.

\subsection{uncertainties}

%
%
Four of the fixed parameters in the adjustment to the FIR, 
($\phi_1, r_{\rm f}, h_{0,{\rm a}}, Z_\sun$), as well as the 
assumed temperature gradient of the disk component, 
are each varied in turn from their
standard values, and the FIR adjustment redone. 
The resulting values of the adjusted parameters
are shown in Table \ref{dustmods}. In all these fits
$R_{\rm s}$, the radius of the local arm, was adjusted to 
within $0.1$ percent of $1R_\sun$, and has thus
been left out of the table. The resulting $\chi^2$ of these
fits fell within 1 percent of that of the adopted standard model,
with the exception of the $\phi_1=3 \deg$ and $Z_\sun=10 \pc$ models
whose $\chi^2$ deviated by less than 3 percent.
The variance of the resulting parameters are used as 
estimates of the uncertainties given in Table \ref{dustpars}.

\placetable{dustmods}

From Table \ref{dustmods} notable correlations between $Z_\sun$ and adjusted
parameters are worth mentioning. The strongest is that between 
$Z_\sun$ and the linear flare parameter $h_1$ of the dust disk,
and we have excluded these models in the calculation of the 
uncertainty of $h_1$.
There are also strong correlations with $Q_{240}$
and the spiral parameters $c_{\rm a}$ and $h_{1,{\rm a}}$.
For the local arm many of the parameters are correlated
to the value of $\phi_1$. In the case of the density and 
emissivity this is expected, however the initial intention of the 
``geometrical'' parameters $\phi_2/\phi_1$, $w_{\rm s}/\phi_1$, 
and $Z_{\rm s}/\phi_1$ was to allow an eventual adjustment of
$\phi_1$ alone in the adjustment to the NIR data 
to arrive at the relative
placement of the local arm. If this were possible then these
geometrical parameters would be independent of the choice of the
initial $\phi_1$. However, this is not realized. For this reason 
$\phi_1$ is kept fixed in the NIR adjustment. 

%
%
To estimate the uncertainties in the stellar parameters the ten 
parameter sets derived 
from the FIR adjustments shown in Table \ref{dustmods} were
in turn used for a suite of NIR adjustments, allowing an 
estimate of the uncertainties in the stellar parameters to be made.
In contrast to the dust model, the stellar model only has one fixed 
parameter, the scale height of the stellar spiral arms, and this parameter
is effectively varied in this suite of adjustments because 
$h^*_{\rm a} = h_{0,\rm a}$.
The resulting sets of parameters are shown in Table \ref{starmods}.

\placetable{starmods}

Inspection of the disk parameters for the different models shows evidence
of clustering about two solutions, probably a result of two minima in 
$\chi^2$ space, one with an average scale length of $2.56 \kpc$ (3 models),
and the other with an average scale length $2.19 \kpc$ 
(for $R_\odot = 8\kpc$). The standard 
model belongs to the latter group and the uncertainties for the disk 
parameters given in Table \ref{starpars} is from the variance of the 
seven models belonging to this shorter radial scale length solution.
The $\chi^2$ only slightly distinguishes between these two solutions,
the longer scale length solutions having an average $\chi^2$ 2 percent 
lower than that of the standard model.
The variance from all ten models is given in parenthesis. The other
parameters do not seem to reflect this clustering, with the exception
of $\kappa_V$. Excluding the three longer scale length models reduces 
the uncertainty of this parameter to 0.0025. Meanwhile the uncertainties
for the dust densities and emissivities are found after excluding
the $\phi_1 = 3 \deg$ and $Z_\sun = 10 \pc$ models, which have 
spiral arm emissivities that are more than three standard deviations
from the mean. Again, the variances from all ten models is given 
in parenthesis. 

The magnitude of the relative uncertainties are in most cases between 10 and
15 percent, though are higher for the 
parameters associated with the spiral arms. 
The data in or near the GP, where most of the information resides
for the spiral arm parameters, makes up a relatively small fraction
of the $\chi^2$, with the result that these parameters
are not as well determined. Possibly another choice of merit function
more sensitive to the GP emission would render a better determination of
these parameters.

\subsection {Alternative models}

Besides the above models we have also 
run a suite of models to test the sensitivity of our results 
with respect to the data selection and specific features in the models.
Adjustments to data restricted to smaller latitudes ($|b| < 20 \deg$),
or with a model lacking a central hole in the dust disk, or with a radial
cutoff in the dust disk imposed at $1.75 R_\odot$, each produced 
estimated parameters that coincided with the standard model
well within the given uncertainties.
However, the adjustment is sensitive to varying the cut in galactic longitude,
as including emission within $20\deg$ of the GC will include emission 
from structures not described by the model, such as the Galactic bulge,
and excluding emission within $30\deg$ of the GC would exclude important
arm tangents needed to contrain the spiral arms. 

Other alternative
stellar models worthy of discussion are given in Table \ref{altmods}.
A model with a flair in the stellar disk, starting at the same radius
as the flair in the dust disk, was fit to the data. The resulting flair 
parameter was quite modest, 6.6 pc/kpc, less than half that found for
the dust disk, but leads to a moderately larger scale height.
No significant improvement of the $\chi^2$ is seen ($< 1$\%), thus no
positive evidence for such a structure can be inferred.
Removing the disk cutoff results in a significant change 
of only a few parameters, while the $\chi^2$ was only slightly 
larger than that of the standard model; this is another feature of 
the stellar disk which is not well constrained. However, the necessity
of the NIR offsets is confirmed. In this case $\chi^2$ increases
more than 17 percent when the offsets are set to zero, 
but most of the parameters are not significantly
different than the normative values. 

A model in which the dust emissivity of the dust spirals was fixed 
to unity was adjusted; though this has a significant effect on the
amplitude of the stellar spirals, other parameters are not affected,
nor is the $\chi^2$ very different from the standard model. This
insensitivity shows that the dust density (and emissivity) associated
with the spiral arms are not well constrained, but likewise neither
are the other parameters affected by this uncertainty.
Another adjustment was done with the scale height of the stellar 
spirals set to twice that of the dust spirals. 
This modification significantly changes some of the parameters, including 
the disk scale length, but is not favored as the total $\chi^2$ 
is 3 percent higher than for the standard value. This and other
experiments confirm that a small scale height is favored for the
stellar spirals.


In order to discern the nonaxisymmetric structure in the Galactic disk
from the NIR emission, four different models were
adjusted against the data, three spiral models and an purely
axisymmetric model.
Figures \ref{spmodj} and \ref{spmodk} show the GP emission profiles 
for longitudes $|l| < 90\deg$ of the four models, and Table \ref{spmods}
gives their parameters and $\chi^2$ in the GP. The two logarithmic 
spiral models use a single amplitude parameter $B$ for both wavebands,
consistent with the hypothesis that the spiral structure has
the same stellar population as the disk. 
Meanwhile the sheared spiral model is formulated under the
assumption that they primarily consist of young stars born from the spiral
arms traced by the dust, and are thus given parameters that 
differ from one waveband to the other. As can be seen in Figures
\ref{spmodj} and \ref{spmodk}, the $m=2$ log spiral model and the sheared 
spiral model perform about equally well, while the nonaxisymmetric
and $m=4$ models fail to produce emission apparently associated with
the spiral arms, particularly in the directions to the tangents
of the Scutum arm, $l \simeq 30$ and $-50\deg$. 

At positive longitudes no clear evidence of 
the Sag-Car arm is present; the axisymmetric and $m=2$ log spiral models 
reproduce the emission profile at $l > 40 \deg$ as well as the other
two spiral models, though the former two models do not possess the
Sag-Car spiral arm. However, a closer comparison of the $m=2$ log and
sheared spiral models suggest that the Sag-Car arm is indeed present,
at least at negative galactic longitudes;
the parameters common to the $m=2$ log and
sheared spiral model are similar, though
the very large pitch angle of the $m=2$ log spiral produces a geometry
significantly different from that of the sheared spiral model (see
Figure \ref{spmap}). 
This high estimate of the pitch angle is most likely due to an effort in 
the model to account for emission at longitudes $l < -50 \deg$, 
which is more prominent in J. Indeed, adjustment of an $m=2$
log spiral model to the K band alone gives a smaller estimate of the 
pitch angle of $15.6 \deg$ \citep{DS00}. 
All four of the models are inadequate in reproducing emission for 
$|l| < 30\deg$.

In fitting the four above models to the NIR data it was necessary to fix
the emissivity of the spiral arms in the dust for the axisymmetric case.
Decoupling of the emissivity and density of the dust spiral component was
successful for the other three models, though questionably reliable for 
the $m=4$ model that gave an emissivity much higher than the disk 
component.


\section{Discussion}

\subsection{axisymmetric structure}

The scale length of the axisymmetric component of the dust distribution
is here found to be $0.28 R_\odot$, which is shorter than that 
of F98 ($0.37 R_\odot$), and significantly
shorter than that of either \citet{SMB97} ($0.48 R_\sun$) or
\citet{davies97} ($0.62 R_\odot$). 
Apparently a more extended dust component is not needed when the distribution 
of the warmer (nonaxisymmetric) components are taken into account.
Other models of the Galactic dust distribution are based on correlating
the dust density with the gas (hydrogen) surface density 
(e.g. \citet{Ortiz93}). Such models
point out that an exponential disk model is not appropriate in the 
central regions of the Galaxy ($r < 0.5 R_\odot$), where the gas 
surface density shows a hole. 
Nevertheless, a characteristic scale length can be assigned to
the outer regions, and these show a much wider range than
those stated above due to uncertainties in the metallicity gradient
and the CO:H$_2$ ratio. The latter is usually taken as constant with
galactocentric radius, but more recently it has been argued that
this ratio may vary considerably with galactocentric radius \citep{sod97}. 

A feature seen in the HI distribution that we have included in our model
is a flair in the disk 
scale height. \citet{Mal95}, using a Gaussian vertical density profile 
for the HI, finds a scale height that flares from $\sim 100\pc$ to 
$\sim 220\pc$ going from 0.5 to 1.0 $R_\sun$. (Here and in what follows,
quantities expressed in parsecs are derived on the assumption that 
$R_\odot = 8\kpc$.)
We find a smaller initial scale height and more modest gradient, 
the scale height increasing from $134\pc$ to $188\pc$ at the Solar Circle.
These values are comparable to the constant scale height of F98
($152\pc$), but smaller than \citet{davies97} much higher value of $470\pc$.
A possible source of bias in our determination of the dust scale height
is the assumption of no vertical temperature gradient, which if present
would result in the scale height of the dust being under estimated.
However, we expect the vertical temperature gradient to be small for the 
dust disk component, as the scale height of the stars heating the dust
is significantly larger than that of the dust.

The isotropic offset parameter $Q_{\rm 240}$ corresponds to the CIB. 
Our value for the CIB, $1.07 \pm 0.15 \mjy$, is
in agreement with other determinations 
($1.17 \pm 0.53 \mjy$, \citet{Fink00}; 
$0.91 \pm 0.15 \mjy$, \citet{Lag99};
$1.09 \pm 0.20 \mjy$, \citet{dirbe1}).
All estimates agree when an additional systematic 
error of $\pm 0.20 \mjy$ is taken into account. 

Our estimate of the stellar radial scale length 
of the Galactic disk ($0.28 R_\odot$) is shorter than
older determinations from NIR data (0.38 $R_\odot$, \citet{Kent91}) .  
We discuss the implications of this result in the conclusions.
Our determination of the stellar and dust scale lengths are most
sensitive to their distribution between .5 and 1 $R_\odot$.
Our model is not sensitive to a cutoff in the dust density, 
though we do find a cutoff in the stellar light distribution at
approximately $10.5\kpc$, which is comparable though smaller than 
that found by \citep{ruphy96} from a starcount analysis of DENIS data.
However, it is still larger than would be expected from the scale length,
as compared to external disk galaxies \citep{Poh00}.


Traditional estimates of scale heights for Galactic stellar populations 
varies from approximately 90 to 390 pc, and is commonly correlated with
absolute magnitudes, stellar type or age in stellar distribution
models built to reproduce starcount data \citep{BAC80,Ortiz93,HRCI}.
Meanwhile luminosity density models which reproduce NIR emission
on the sky, like our own presented here, have reported scale heights
of 247 pc \citep{Kent91}, 276 pc \citep{SMB97} and 334 pc (F98).
Our estimate of 282 pc for the scale height falls amoung these
values. However, scale height estimates do not lend themselves to 
direct comparison as scale lengths do, because the formulation of the 
vertical density profile does not enjoy a universal consensus as does
the radial profile. In the past vertical exponential profiles
were most common, even after it was shown that a $\sech^2$ variation
describes an isothermal population in the infinite plane approximation.
Recently this situation has changed with more sophisticated models
that employ dynamical constraints that have called into question 
the traditional exponential vertical profiles \citep{HRCI, HRCII}. 
Such models do not possess a simple vertical profile, but only approach
an exponential far from the GP, similar to a $\sech^2$ profile.
However, this common behavior among the various proposed
profiles does not assist us in making comparisons via asymptotic 
scale heights, as the integrated luminosity is mainly determined
by the density variation near the GP.

We note at this point in the discussion two possible sources of biases
in our determination of scale heights. The first is the assumption of
a constant dust temperature with respect to $z$ and it's possible
affect on the estimation of the dust scale height, mentioned above,
which if present would cause an underestimation of the stellar scale height.
The other source of 
potential bias is the removal of luminosity from point source removal, 
which preferentially takes place at higher galactic latitudes where the 
median background light from unresolved stars is lower. This effect could 
lead to an underestimate of the vertical scale height. The isotropic term 
in the emission model may correct for this to some degree, but in the 
future it will be desirable to make a more accurate correction for 
this effect.

\subsection{nonaxisymmetric structure}

Similar to the results of Freudenreich \citep{fred96, fre98} we find
a warp that starts within the Solar Circle with larger amplitudes 
in the dust than in the stars. At $r = 10\kpc$ our stellar warp has an
amplitude of $0.25\kpc$ ($R_\odot = 8\kpc$), 
while F98 finds $0.18\kpc$ (his Model S). 
From FIR emission we find a dust warp 2.7 times larger, 
whereas F98 finds an amplitude that is 1.8 times larger as
inferred from NIR dust absorption; his dust warp amplitude agrees well
with the estimated amplitude of 0.3 to 0.4 kpc that has been given for 
the HI \citep{BUR94}. Our warp amplitude in the dust does agree, however, 
with the warp inferred from the OB stellar distribution \citep{SMA98}.
In the NIR the most important effect of a warp that starts within 
the Solar Circle is to induce a local tilt 
($\tan \theta = h_{\rm w}(R_\sun)/R_\sun$), 
as evidenced in the NIR emission profiles at low galactic 
latitudes (Section 4.2). For the stars we find a local tilt of 
$\theta = 0.2 \deg$
while Freudenreich's warp renders a local tilt of $\theta = 0.5 \deg$. 
A tilt of the stellar distribution with respect to the conventionally 
defined $b=0$ plane has also been noted in NIR data by \citet{Hamm95}, 
though they propose that a global tilt of the 
entire stellar disk with respect to the $b=0$ plane is responsible for
this feature. The motivation for this alternative model is that 
radio observations seem to indicate the Galactic warp starts beyond the Solar 
Circle, though this is not well contrained by the radio data.
Evidence for a warp being present in the stellar disk is also found in
local stellar kinematics, as seen in Hipparcos data,
though it is more consistent with a warp starting at or beyond the Solar 
Circle \citep{dehnen98}. While the warp both here and in F98 
starts within the Solar Circle, this is not a commonly accepted
feature of the warp. In any case  
our warp starts at a significantly larger galactocentric radius 
($\sim 0.85 R_\odot$) than that of F98 (between 0.5 and $0.56 R_\odot$), 
though this difference may only be a consequence of the different functions 
used to describe the warp amplitude. 
%
%
%

%
With regards to the spiral arms, it was found 
that an adopted map of HII
regions was sufficient to describe the location of peaks in the 
FIR emission features associated
with the spiral arm tangents, though it was necessary to introduce a
reduction factor on the Sag-Car arm. This assumed geometry for the spiral arms
is consistent with a four arm model of the Galaxy, with a pitch angle of
approximately $12.5\deg$, consistent with other spiral tracers,
radio and pulsar data \citep{val95}. We point out that 
the dust density associated with the arms is not well constrained, as
strong absorbtion features from the spiral arms are not present in the NIR 
to assist in the decomposition of the FIR flux density. Furthermore,
there may be significant temperature gradients associated with the arms
as the dust here may be primarily heated by young OB stars. By using a 
single emissivity the widths and scaleheights of this component may be 
significantly underestimated. However, for the same reason that the dust
density in the arms are not well determined, this uncertainty does not 
strongly affect the estimation of the stellar distribution parameters.

Evidence for the spiral structure in the NIR is less evident due to 
extinction, but important
in and near the GP. Three different geometries were
attempted, the geometrical parameters of each model being constrained with 
the NIR data. Our standard model is a sheared version of the 
same four arm spiral model used for the dust. 
Spiral models constructed to describe the distribution of NIR point sources 
have also utilized four arm models, though of a logarithmic form
\citep{Sky92,Ortiz93}. One important difference
between the spiral model presented here and these previous models
is the reduction factor on the size of the Sag-Car arm. The
necessity of such a reduction factor is demonstrated by the 
failure of the $m=4$ logarithmic model in which all four arms are 
treated equally, and it's effect is consistent with the conclusion that
two arms dominate in the NIR. This interpretation is further supported
by the comparable success of a purely two arm spiral model.

We find that the spirals are stronger in K than in J, which also agrees
with observations of other spiral galaxies in these wavebands \citep{Grau98}.
The amplitude of our spirals, however, are smaller than those seen in most
spiral galaxies;  \citet{Rix95} report fractional azimuthal 
variations in the K surface 
brightness, $(\Sigma_{\rm max} - \Sigma_{\rm min})/\Sigma_{\rm min}$,
to be of the order unity for their sample of spiral galaxies, while we 
find here that this quantity is equal to
$(\sqrt \pi h^*_{\rm a}/2 h_*) B_{\rm K} = .32$.  
Also, the fraction of total light in K from the spirals, approximately 10\%,
is smaller than that seen in most other galaxies \citep{Seig98b}.
Our weaker spiral arms may be due to their scale 
height being underestimated relative to that of the disk. 
Such an underestimate could be due to K supergiants 
dominating the spiral flux density in the GP, or be a 
consequence of underestimating the dust column density associated with 
the arms. 
Our spiral model also renders an estimate of the corotation radius 
of the Milky Way, which is smaller than other estimates, but which
gives a ratio $R_C/r_* = 2.9$ consistent with determinations of other
spiral galaxies \citep{Gros98}.

It is interesting to note that our spiral model 
is consistent with the dynamical model
of the Milky Way's spiral arms constructed by \citet{AL97}, after a trivial 
$90\deg$ rotation, which has four arms but with two arms dominating.
Additional evidence of spiral arms comes from recent 
results on the star formation history of the solar neighborhood 
\citep{Rocha00a,Hern00}, that indicate an intermittent or episodic star 
formation rate for the Galactic disk. A periodicity of $\sim .5$ Gyr is
suggested by \citet{Hern00}, and they point out that this periodicity is 
consistent with crossing a two arm spiral.
Assuming $\Omega_o = 25 \kmskpc$, our estimate for the corotation radius,
$R_C = 0.83 \pm 0.05 R_\odot$ gives a range of star formation periodicity
from 0.5 to 0.9 Gyr for a two arm geometry. 

\section{Summary}

We fit joint models for the Galactic dust and stellar distributions
to the COBE FIR and NIR emission.  Our dust model has 22 parameters
adjusted to fit 173,569 data points, while our parameteric
model of the stellar emission has 19 adjusted parameters that fits the 
large scale NIR emission features in 304,742 data points.

Our model of the Milky Way has several intriguing results:

(1) We find a small scale length for the stars in the Milky Way disk.
Our estimate of the stellar radial scale length 
of the Galactic disk ($0.28 R_\odot$) is shorter than
older determinations from NIR data (0.38 $R_\odot$, \citet{Kent91}) .  
It is, however, consistent with recent studies using 
NIR data, such as IRAS point sources (0.33 $R_\odot$, \citet{Ortiz93}), 
DENIS data (0.27 $R_\odot$, \citet{ruphy96}),  
the Two Micron Galactic Survey (0.25 $R_\odot$, \citet{PGJ98}), 
and earlier determinations based on the DIRBE data 
(0.35 $R_\odot$, \citet{SMB97}; 0.31 $R_\odot$, F98).
It also agrees with analysis of local stellar kinematics
(0.29 -- 0.36 $R_\odot$, \citet{fux94}; 0.2 -- 0.34 $R_\odot$, \citet{BS97})
including recent use of Hipparcos data (0.29 -- 0.33 $R_\odot$, \citet{DB98b}; 
however 0.19 -- 0.25 $R_\odot$ in \citet{Bien99}). 
Recent studies at visual wavelengths 
are beginning to converge on a shorter scale length as well
(0.21 -- 0.36 $R_\sun$, \citet{Ojha96}, 0.36 $R_\sun$, \citet{Gould97}), 
though it is yet far from unanimous
(0.47 -- 0.94 $R_\sun$, \citet{Mendez98}; 0.43 -- 0.56 $R_\sun$, \citet{Ng95}).
We note that \citet{jong96a} finds that the NIR scale length 
is 20 percent smaller than the optical scale length in external galaxies.  He
argues that this is the signature of inside-out galaxy formation:
the outer regions of spiral galaxies are younger than the inner regions.
This small scale length has a number of important implications
for the Galactic mass distribution: ({\em i}) it implies that
maximal disk models are a good fit to the Galactic rotation curve
\citep{DB98a} and implies a low central density for the dark matter halo which
contradicts CDM simulations\citep{sellwood}; ({\em ii}) can 
increase the microlensing optical depth towards the Galactic bulge 
and will produce more long duration events \citep{Bin00, Sack97};
({\em iii}) it implies that
the Galaxy is significantly smaller than our neighbor M31 \citep{Hiro83}.

(2) We find that a two arm spiral structure 
dominates the NIR nonaxisymmetric emission, supporting the earlier 
simple analysis of Drimmel (2000).  What stellar population
dominates our spiral arms in the near-IR?  
Diffuse NIR emission associated
with the spiral arms will have contributions from both young
stellar populations, such as K supergiants, and from the old disk 
population if a spiral density wave is present.
The geometry from these two types of populations will not necessarily
be the same, as new stars are born from gas subject to hydrodynamic forces.
Our standard model of sheared HII arms, that are redder than the disk,  
is consistent with the
assumption of the spiral emission being produced by young K-supergiants,
while a $m=2$ logarithmic spiral model with a more 
open structure and the same color as the disk
is consistent with a density wave. 
The success of these two geometries may indicate a need
to include both types of spiral emission. The
small scale height of the arms favors the hypothesis of 
a young population dominating the spiral emission, though the vertical 
profile of a density wave perturbation in the stars is not known.
In external galaxies with active star formation, young stars dominate 
the spiral arms even in K band, while
in other galaxies the K band light is tracing the older stars
and the stellar mass \citep{Rho98}.  Once the 2MASS data is available
for our spiral arms, we will be able to address this question in
our Galaxy.

(3) The Galactic warp is here found to start within the Solar Circle
and to have different amplitudes  
for the dust and the stars. If this is indeed the case it is an
important clue as to the nature of the warping 
mechanism, suggesting that hydrodynamic or
magnetohydrodynamic forces are important. It may also suggest that the 
Galactic warp is not a long-lived feature. One 
mechanism that would cause a short-lived warp is an interaction with
one or more companions; this could cause a different response in 
the gas and stars of the galaxy. Both the Magellanic Clouds 
\citep{wein95, wein00} 
and the Sagittarius Dwarf \citep{ibata98aa} have been 
suggested in this context.  

However, there is an important caveat to this last result, namely that 
the {\em only} large-scale vertical distortion present in the Galactic disk
is in the form of a warp. That this may be inadequate
is suggested by several points. First, the amplitude of the
warp in the dust is apparently inconsistent with radio data of HI. 
Secondly, the phase of the warp is found here to be less than
$1\deg$, a result also found by F98.
Assuming that the location with respect to the Galactic warp is not
a determining factor for the existence of observers, the probability of 
our being this close to the line-of-nodes ($2 \phi_{\rm w}/ \pi$) is less 
than 1 in 100. The minimizing of $\phi_{\rm w}$ effectively maximizes the 
local tilt for a fixed warp amplitude; the presence of other distortions
could produce additional tilt in the dust than what would be produced
by a warp alone. In addition 
there are several features to the adjustment of related parameters
which may be pointing to an inadequacy, namely the inability to 
adjust $Z_\odot$ in the initial FIR fit and the need for
a displacement specific to the local dust feature.

These points suggest the alternative interpretation that, in addition 
to a global warp, there 
are small-amplitude oscillations affecting the local structure
of the gas and dust, that are not described in the model.
Evidence for vertical displacements within the Solar Circle
can be found in the COBE data, particularly with regards to residuals 
at the spiral arm tangents. 
\citet{Sky99} has also noted evidence that the spiral arms
show displacements out of the GP in NIR point source data.
It has been noted in numerical experiments that small amplitude vertical 
displacements in the disk could result from oscillations excited by 
Galactic satellites \citep{Edel97}, though displacements specific
to the spiral arms would have to be of a hydrodynamic nature.

We are entering a golden age for Galactic astronomy.  Near-infrared
and far-infrared observations are revealing structure that is 
hidden in the optical. Our analysis here shows that there is a wealth 
of Galactic structure information
in COBE two dimensional data.  This data will soon be complemented
by a 2MASS inventory of the bright stars in our galaxies.  When this
structural data is combined with information from the coming generation
of astrometric satellites (FAME, SIM and GAIA), we will finally
have a dynamically detailed picture of our home, the Milky Way.


\acknowledgments

Thanks are extended to Janet Weiland for assistance in installing 
the {\small UIDL} COBE data analysis package.
The COBE datasets were developed by the NASA
Goddard Space Flight Center under the guidance of the COBE Science
Working Group and were provided by the NSSDC.

\clearpage

\figcaption[figs/jerror.ps]{
The local error in the J passband, with respect to galactic latitude, 
evaluated at each pixel location within a 7$\times$7 pixel window. 
\label{jerror}}

\figcaption[figs/fdat.ps]{
Map of the ratio of the smoothed to unsmoothed data (top), and 
the estimated local error $\sigma_{240}$ (bottom). 
The maps are Mollweide galactic projections
for $|b| < 30\deg$, on logarithmic scales, indicated by the grey scale
bars. Blanked pixels are regions excluded from the present analysis (see text).
Positive galactic longitudes are to the left of center.
\label{fdat}}

\figcaption[figs/emprof.ps]{
The relative emissivity of the dust disk as a function of temperature
galactocentric radius.
\label{emprof}}

\figcaption[figs/fmaps.ps]{
$240 \micron$ skymaps of the smoothed DIRBE data (top), 
the modeled emission (middle), 
and the relative difference 
between the modeled and observed emission (bottom). The top two maps
are on a logarithmic scale, while the bottom map is on a linear scale.
 \label{fmaps}}

\figcaption[figs/firgp.ps]{
The $240\micron$ emission profile
for the data (X's) and model (diamonds) within $0.17 \deg$ of
the GP ($b=0$), on a logarithmic scale ($\log D$).
 \label{firgp}}

\figcaption[figs/dprof.ps]{
$240\micron$ emission profiles within $0.17 \deg$ of the
indicated galactic latitudes. Scale and symbols are as in previous figure.
 \label{dprof}}

\figcaption[figs/latdev.ps]{
The logarithm of the absolute FIR deviations, between the data and the model,
as a function of galactic latitude.
 \label{latdev}}

\figcaption[figs/threej.ps]{
Skymaps of the observed J band emission (top), 
the predicted J band emission from the 
stellar distribution model (middle), and the relative difference map
between model and data (bottom). Black pixels in the upper two maps
show pixels removed from the data set, due to blanking out selected
regions, such as the Galactic center, or rejected point sources.
Upper two maps are on a log scale, while the bottom is on a linear scale.
 \label{threej}}

\figcaption[figs/threek.ps]{
Same as previous figure, but for the K band.
 \label{threek}}

\figcaption[figs/nirgp.ps]{
Observed (X's) and modeled (solid line) J and K band emission within
0.2 deg of the GP for the sheared spiral arms model.    
Vertical scale is in magnitude units ($-2.5 \log D_{\rm b}$).
 \label{nirgp}}

\figcaption[figs/jprof.ps]{
J band emission as observed and modeled within $0.2 \deg$
of the indicated latitudes. Scale, symbols and units are as in previous figure.
 \label{jprof}}

\figcaption[figs/kprof.ps]{
K band emission as observed and modeled within $0.2 \deg$
of the indicated latitudes. Scale, symbols and units are as in previous figure.
 \label{kprof}}

\figcaption[figs/dmap.ps]{
Surface density map of the dust, as inferred from the dust density model.
Small black dot (upper center) shows the position of the Sun, which 
nearly lies on a small local feature, known as the Orion arm.
Arms are incomplete on the side opposite the Sun due to incomplete HII data.
 \label{dmap}}
\figcaption[figs/smap.ps]{
Surface brightness map of the Milky Way in the K band. Bright dot (upper
center) indicates the position of the Sun.
 \label{smap}}

\figcaption[figs/scaleh.ps]{
Scale heights as a function of radius for the stellar disk (dashed line),
the dust disk (solid line), and spiral arms (dot dash line), assuming 
$R_\sun = 8 \kpc$.
 \label{scaleh}}
\figcaption[figs/intarm.ps]{
Arm - interarm density ratio for the stellar (dashed curve) and the 
dust spiral arms (solid curve).
 \label{intarm}}

\figcaption[figs/spmodj.ps]{
Predicted J band GP emission profiles at $|l| < 90 \deg$ 
for an axisymmetric model and three spiral models,
$m=2$ and $m=4$ logarithmic spiral models, and a sheared HII spiral model,
plotted against the data (X's).
 \label{spmodj}}
\figcaption[figs/spmodk.ps]{
Same as previous, but for the K band.
 \label{spmodk}}
\figcaption[figs/spmap.ps]{
Schematic of the Galaxy showing the four spiral arms as mapped by HII regions
and the dust (bold lines), the sheared arms in the K band (stars), and
the arms in the two-arm logarythmic model for J and K band fit (dashed) and
the K band fit alone (solid) \citep{DS00}. The HII spirals are incomplete 
on the opposite side of the Galaxy due to lack of data.
 \label{spmap}}

%
%

\clearpage
\singlespace


%



\begin{deluxetable}{llll}
\tabletypesize{\footnotesize}
\tablecaption{Dust parameters. \label{dustpars}}	
\tablewidth{0pt}
\tablehead{
\colhead{Parameter} & \colhead{Symbol}   & \colhead{Value}   & \colhead{Uncertainty}
}
\startdata
Disk component:              &           	&                 &    \\
density                      & $\rho_0$  	& 1098 MJy/sr/kpc &  41  \\
scale length                 & $h_r$     	& 2.26 kpc        & 0.16  \\
base scale height            & $h_{0,{\rm d}}$ 	& 134.4  pc       & 8.5   \\
linear flair coefficient     & $h_{1,{\rm d}}$ 	& 14.8 pc/kpc     & 9.9   \\
flair radius\tablenotemark{\dagger} & $r_{\rm f}$ & 4.40   kpc    & -- \\
$240\micron$ offset          & $Q_{240}$ 	& 1.07 MJy/sr     & 0.15   \\
                             &           	& 	          &    \\
\tableline
Local Arm:           & 		          	&       	&    \\
Radius at $\phi=0$   & $R_{\rm s}$              & 8.001 kpc     & 0.001   \\
pitch angle          & $p_{\rm s}$              & 7.33 deg      & 0.45   \\
gap parameters:
		     & $\phi_1$\tablenotemark{\dagger} & 1.500 deg  & --   \\
                     & $\phi_2/\phi_1$          & -0.643            & 0.017 \\
half width           & $w_{\rm s}/\phi_1$       & 16.28 pc/deg      & 0.70   \\
height               & $Z_{\rm s}/\phi_1$       & 11.9  pc/deg      & 2.9   \\
central density      & $\rho_{\rm s}$           & 156 MJy/sr/kpc    & 29   \\
emissivities         & $k_{+}/k_{-}$            & 3.09              & 0.52   \\
                             &                  &                   &    \\
\tableline
Spiral arms:           &           &        &    \\
emissivity\tablenotemark{\dagger} & $k_{\rm a}$	 & 1.00            & --   \\
density                      & $\rho_{\rm a}$    & 162 MJy/sr/kpc  & 28   \\
cut-off radius               & $r_{\rm m}$       & 6.71 kpc        & 0.48   \\
cut-off scale length         & $r_{\rm a}$       & 2.07 kpc        & 0.22   \\
arm width coefficient        & $c_{\rm a}$       & 64.1 pc/kpc     & 8.8   \\
initial scale height\tablenotemark{\dagger} 
		             & $h_{0,{\rm a}}$   & 80.0 pc       & --   \\
flair radius                 & $r_{\rm f,a}$     & 5.48 kpc      & 0.13   \\
flair coefficient            & $h_{1,{\rm a}}$   & 14.6 pc/kpc$^2$  & 1.4 \\
Sag-Car reduction factor     & $f_{\rm r}$   	 & 0.407         & 0.036   \\
                             &           &               &    \\
\tableline
Warp:       			&           &       &    \\
Radius that warp starts   & $R_{\rm w}$     & 6.993 kpc          & 0.046 \\
Phase of the warp         & $\phi_{\rm w}$  & -0.12 deg          & 0.45  \\
amplitude coefficient     & $a_{\rm w}$     & 72.8  pc/kpc$^2$   & 7.0  \\

\enddata
\tablenotetext{\dagger}{Fixed parameters.}
\end{deluxetable}

\clearpage

%
%
\begin{deluxetable}{llll}
\tabletypesize{\footnotesize}
\tablewidth{0pc}
\tablecaption{Stellar parameters. \label{starpars}}
\tablehead{
\colhead{Parameter} & \colhead{Symbol}   & \colhead{Value}   & \colhead{Uncertainty}
}
\startdata
Disk:         &  & &    \\
normalization in J &	$\eta^0_{\rm J}$ & 14.7 MJy/sr/kpc &  2.3  (4.2) \\
normalization in K &	$\eta^0_{\rm K}$ & 11.6 MJy/sr/kpc &  1.7  (3.2) \\
radial scale length&	$r_*$ 	         & 2.264 kpc	&  0.083 (0.19) \\
scale height	&	$h_*$ 	         & 282.2 pc 	&  7.9  (20.)\\
cutoff radius	&	$r_c$            & 10.52 kpc	&  0.34  \\
                & & &   \\ 
\tableline
Spiral arms:    & & &   \\
amplitude in J     & $B_{\rm J}$                  & 0.86	& 0.25 \\
amplitude in K     & $B_{\rm K}$                  & 1.28	& 0.24  \\
mean age in J	   & $\tau_J$                     & 5.5 Myr   	& 3.6  \\
mean age in K	   & $\tau_K$                     & 17.6 Myr  	& 3.8 \\
arm width coeff.   & $c_{\rm J}^*$                & 142. pc/kpc	& 43. \\
arm width coeff.   & $c_{\rm K}^*$                & 143. pc/kpc	& 69.  \\
corotation radius  & $R_{\rm C}$                  & 6.66 kpc	& 0.39 \\
                        &			& &  \\
\tableline
Dust:             &      	&  &   \\
V opacity 	&  $\kappa_{\rm V}$  &  0.0180 (MJy/sr)$^{-1}$ 	& 0.0029 \\
emissivities 	& $k_{+}$           &  3.98  		& 0.35 (0.45) \\
		& $k_{-}$           &  1.29 		& 0.10 (0.26) \\
	        & $k_{\rm a}$       &  2.07	 	& 0.37 (2.9) \\
densities 	& $\rho_{\rm s}$    &  121. MJy/sr/kpc	& 23.  (32.) \\
	        & $\rho_{\rm a}$    &  61. MJy/sr/kpc 	& 16.  (26.) \\
                & & &    \\ 
\tableline
Miscellaneous:   & &  &   \\
height of Sun 	&  $Z_\sun$          & 14.6 pc			& 2.3  \\
warp coefficient & $a_w$   	     & 27.4 ${\rm pc/kpc}^2$	& 2.5 \\
J offset	&  $Q_{J}$ 	     & -0.0684 MJy/sr		& 0.0082 \\
K offset	&  $Q_{K}$ 	     & -0.0744 MJy/sr 		& 0.0069 \\

\enddata

\end{deluxetable}

\clearpage
%
%
\begin{deluxetable}{clcccccccccc}
\tabletypesize{\scriptsize}
\rotate
\tablewidth{0pc}
\tablecaption{Dust models with variable constants. \label{dustmods}}
\tablehead{
\colhead{} & \colhead{} &
\colhead{$h_{0,{\rm a}}=$} & \colhead{} &
\colhead{$r_{\rm f,a}=$} & \colhead{} &
\colhead{$\phi_1=$} & \colhead{} &
\colhead{$Z_\odot=$} & \colhead{} &
\colhead{$\nabla T=$} & \colhead{} 
\\
\colhead{Parameter} & \colhead{Units}   &
\colhead{60 pc,} & \colhead{100 pc} &
\colhead{$0.5 R_\sun$,} & \colhead{$0.6 R_\sun$} &
\colhead{$1.0\deg$,} & \colhead{$3.0\deg$} &
\colhead{10 pc,}   & \colhead{20 pc} &
\colhead{$-5.5$K$ /R_\sun$,}  & \colhead{$-7.5$K$ /R_\sun$}
}
\startdata
multicolumn{1}{l}{Disk:}             	& & & & & & & & & & & \\
 $\rho_0$ & MJy/sr/kpc	 & 1122 & 1112 & 1057 & 1096 & 
	1120 & 1172 & 1018 & 1060 & 1084 & 1067 \\
 $h_r$  & kpc     	 & 2.21   & 2.28   & 2.26   & 2.31   & 
	2.19 & 1.93 & 2.52 & 1.99 & 2.22 & 2.32 \\
 $h_{0,{\rm d}}$ & pc	 & 139.0   & 131.5   & 132.4   & 135.9   & 
	135.3 & 157.3 & 135.0 & 153.1 & 136.7 & 134.5 \\
 $h_{1,{\rm d}}$ & pc/kpc & 16.5    & 14.7    & 15.3    & 12.3    & 
	20.9  & 44.1  & 0.8 & 46.4 & 13.5 &  14.5 \\
 $Q_{240}$ & MJy/sr & 1.04   & 1.16   & 1.07   & 1.13   & 
	1.00 & 0.97 & 1.42 & 0.82 & 1.10 & 1.09 \\
                  	& & & & & & & & & & & \\ 
\tableline
multicolumn{1}{l}{Local Arm:}      &  & & & & & & & & & &\\
 $p_{\rm s}$ & deg      & 7.35  & 7.36  & 6.88  & 7.38  & 
	7.96  & 6.37 & 7.74 & 7.88 & 7.17 & 7.21 \\
 $\phi_2/\phi_1$     &  & -0.648 & -0.635 & -0.634 & -0.660 & 
	-0.654 & -0.640 & -0.666 & -0.619 & -0.680 & -0.657 \\
 $w_{\rm s}/\phi_1$ & pc/deg & 16.43   & 16.25   & 16.65   & 16.21   & 
	16.88   & 18.46   & 15.85   & 17.08 & 17.09 & 16.50 \\
 $Z_{\rm s}/\phi_1$ & pc/deg & 12.0   & 12.1   & 11.8   & 11.9   & 
	17.5   & 5.7    & 9.5    & 14.7 & 11.8 & 11.8 \\
 $\rho_{\rm s}$ & MJy/sr/kpc & 161 & 158 & 160 & 155 & 
	223 & 101 & 152 & 183 & 152 & 158 \\
 $k_{+}/k_{-}$       &  & 2.84  & 2.82  & 2.71  & 3.15  & 
	2.90  & 1.47 & 3.31 & 2.09 & 2.76 & 2.86 \\
                & & & & & & & & & & & \\
\tableline
multicolumn{1}{l}{Spiral arms:}    &  & & & & & & & & & & \\
 $\rho_{\rm a}$ & MJy/sr/kpc    & 174 & 130 & 165 & 160 & 
	148 & 87 & 139 & 104 & 164 & 162 \\
 $r_{\rm m}$ & kpc            	& 6.36  & 7.45  & 6.56  & 6.77  & 
	7.16  & 7.85 & 7.19 & 7.53 & 6.55 & 6.73 \\
 $r_{\rm a}$ & kpc            	& 2.23  & 1.68  & 2.15  & 2.02  & 
	1.81  & 1.59 & 1.78 & 1.74 &  2.14 & 2.05 \\
 $c_{\rm a}$ & pc/kpc         	& 69.6   & 57.4   & 66.3   & 60.7   & 
	67.6   & 81.4  & 53.1 & 81.5 & 64.9 & 64.1 \\
 $r_{\rm f,a}$ & kpc        	& 5.38  & 5.71  & 5.45  & 5.46  & 
	5.69  & 5.38 & 5.32 & 5.34 & 5.42 & 5.48 \\
 $h_{1,{\rm a}}$ &pc/kpc$^2$ & 14.7 & 14.5   & 14.2   & 14.9   & 
	14.6   & 11.3  & 15.3 & 11.0 & 14.2 & 14.5 \\
 $f_{\rm r}$            	& & 0.432 & 0.384 & 0.406 & 0.404 & 
	0.409 & 0.518 & 0.420 & 0.440 & 0.404 & 0.406 \\
                        & & & & & & & & & & & \\
\tableline
multicolumn{1}{l}{Warp: }                  & & & & & & & & & & & \\
 $R_{\rm w}$    & kpc             & 7.016 & 6.973  & 6.990  & 7.000  & 
	7.017 & 7.053 & 7.132 & 7.041 & 6.966 & 6.996 \\
 $\phi_{\rm w}$ & deg             & 0.18 & 0.05 & 0.02 & -0.30 & 
	0.43 & 1.28 & 0.44 & 0.57 & -0.14 & -0.11 \\
 $a_{\rm w}$    & pc/kpc$^2$  & 75.9  & 73.6   & 72.7   & 73.8   & 
	77.5  & 86.7  & 91.8 & 84.4 & 70.3 & 73.4 \\
\enddata
\end{deluxetable}

\clearpage 

%
%
\begin{deluxetable}{clcccccccccc}
\tabletypesize{\scriptsize}
\rotate
\tablewidth{0pc}
\tablecaption{Alternative Stellar models. \label{starmods}}
\tablehead{
\colhead{} & \colhead{} &
\colhead{$h_{0,{\rm a}}=$} & \colhead{} &
\colhead{$r_{\rm f,a}=$} & \colhead{} &
\colhead{$\phi_1=$} & \colhead{} &
\colhead{$Z_\odot=$} & \colhead{} &
\colhead{$\nabla T=$} & \colhead{} 
\\
\colhead{Parameter} & \colhead{units} &
\colhead{60 pc,} & \colhead{100 pc} &
\colhead{$0.5 R_\sun$,} & \colhead{$0.6 R_\sun$} &
\colhead{$1.0\deg$,} & \colhead{$3.0\deg$} &
\colhead{10 pc,}   & \colhead{20 pc} &
\colhead{$-5.5$K$ /R_\sun$,}  & \colhead{$-7.5$K$ /R_\sun$}
}
\startdata
multicolumn{1}{l}{Disk:}         &  & & & & & & & & & &  \\ 
$\eta^0_{\rm J}$ & MJy/sr/kpc & 18.6 & 7.6 & 14.6 & 20.5 & 
8.6 & 15.6 & 14.7 & 16.3 & 14.6 & 9.4 \\
$\eta^0_{\rm K}$ & MJy/sr/kpc & 14.7 & 6.2 & 11.5 & 15.8 & 
6.9 & 12.4 & 11.5 & 13.1 & 11.6 & 7.5 \\
$r_*$ & kpc      & 2.063 & 2.596  & 2.264   & 2.101 & 
2.564 & 2.220 & 2.264 & 2.186 & 2.264 & 2.514 \\
$h_*$ & pc       & 290.5  & 329.5   & 283.3    & 265.1  & 
318.0 & 283.6 & 279.3 & 284.7 & 283.2 & 309.4 \\
$r_c$ & kpc      & 11.48 & 10.66 & 10.55 & 10.45 & 
10.58 & 10.17 & 10.36 & 10.51 & 10.59 & 10.56 \\
        &  & & & & & & & & & &  \\ 
\tableline
multicolumn{1}{l}{Spiral arms:}         &  & & & & & & & & & &  \\ 
$B_{\rm J}$  &                & 1.50 & 0.89 & 0.85 & 0.76 & 
0.94 & 0.66 & 0.88 & 0.58 & 0.82 & 1.01 \\
$B_{\rm K}$  &                & 1.24 & 1.43 & 1.30 & 1.07 & 
1.62 & 1.04 & 1.15 & 1.76 & 1.29 & 1.49 \\
$\tau_J$ & Myr                & 12.7 & 11.4 & 9.0 & 7.1 & 
7.4 & 16.3 & 12.5 & 9.2 & 3.7 & 7.3 \\
$\tau_K$ & Myr                & 24.5 & 13.7 & 17.2 & 23.2 & 
13.7 & 20.4 & 19.0 & 17.9 & 14.9 & 15.4 \\
$c_{\rm J}^*$ & pc/kpc        & 158.  & 198.  & 135.  & 143. & 
145. & 79. & 208. & 72. & 132. & 158. \\
$c_{\rm K}^*$ & pc/kpc        & 310.  & 193.  & 139.  & 153. &
136. & 100. & 206. & 54. & 134. & 165. \\
$R_{\rm C}$ & kpc             & 5.67 & 6.81 & 6.69 & 6.12 &
6.72 & 6.44 & 6.33 & 6.86 & 6.95 & 6.62 \\
                    &  & & & & & & & & & &  \\ 
\tableline
multicolumn{1}{l}{Dust:}               &  & & & & & & & & & &  \\ 
$k_{+}$ &      	  	& 4.16  & 3.46  & 3.64 & 4.36 &
3.53 & 2.64 & 3.94 & 4.03 & 3.62 & 3.42 \\
$k_{-}$ &     	  	& 1.46  & 1.23  & 1.34 & 1.38 &
1.22 & 1.80 & 1.19 & 1.93 & 1.31 & 1.19 \\
$k_{\rm a}$  &    	& 1.77  & 2.46  & 2.80 & 2.37 &
2.94 & 10.65 & 2.65 & 7.99 & 2.84 & 2.56 \\
$\rho_{\rm s} $ & MJy/sr/kpc & 110. & 128. & 119. & 112. &
183. & 56. & 128. & 95. & 116. & 132. \\
$\rho_{\rm a}$  & MJy/sr/kpc & 98. & 53.  & 59.  & 68. &
50. & 8. & 53. & 13. & 58. & 63. \\
                    &  & & & & & & & & & &  \\
\tableline
multicolumn{1}{l}{Misc.:}      &  & & & & & & & & & &  \\ 
$\kappa_{\rm V}$ & (MJy/sr)$^{-1}$ & 0.0177 & 0.0149 & 0.0182 & 0.0194 &
0.0152 & 0.0235 & 0.0170 & 0.0225 & 0.0182 & 0.0155 \\
$Q_{J}$ & MJy/sr      & -0.0441 & -0.0704 & -0.0682  & -0.0680 &
-0.0691 & -0.0672 & -0.0715 & -0.0608 & -0.0673 & -0.0711 \\
$Q_{K}$ & MJy/sr      & -0.0565 & -0.0781 & -0.0743  & -0.0715 &
-0.0791 & -0.0737 & -0.0762 & -0.0669 & -0.0738 & -0.0790 \\
$Z_\sun$ & pc            & 14.9    & 14.4    & 14.8  & 14.9  &
14.8 & 14.8 & 10.0 & 19.8 & 14.8 &  14.5 \\
$a_w$ & ${\rm pc/kpc}^2$ & 25.0  & 25.4   & 26.9    & 29.8  &
26.1 & 27.6 & 33.3 & 26.5 & 26.2 & 26.5 \\
\enddata
\end{deluxetable}

\clearpage

%
%
\begin{deluxetable}{ccccccc}
\tabletypesize{\footnotesize}
\tablewidth{0pc}
\tablecaption{Alternative Stellar models. \label{altmods}}
\tablehead{
\colhead{Parameter} & \colhead{units} &
\colhead{stellar flair} & \colhead{$r_c = \infty$} &
\colhead{$Q_{\rm J,K} = 0.$} &
\colhead{$k_a = 1$} & 
\colhead{$h^*_{\rm a}=2h_{0, \rm a}$} 
}

\startdata
\multicolumn{1}{l}{Disk:}         & & & & & &  \\ 
$\eta^0_{\rm J}$ & MJy/sr/kpc & 12.2 & 13.2 & 10.2  & 14.1 &  22.8 \\
$\eta^0_{\rm K}$ & MJy/sr/kpc & 9.7  & 10.3 &  7.4 & 11.1 &  16.9 \\
$r_*$ 		 & kpc       & 2.319   & 2.255 & 2.302 & 2.273 &  2.06 \\
$h_*$ 		 & pc        & 292.3   & 309.0 & 306.30 & 285.6 & 253.2 \\
$r_c$ 		 & kpc      & 11.10 & $\infty$  & 12.41 & 10.54 &    10.90 \\
         & & & & & &   \\ 
\tableline
\multicolumn{1}{l}{Spiral arms:}         & & & & & &  \\ 
$B_{\rm J}$  	&     	   & 1.06 & 0.97 & 0.69 & 1.33 & 0.14 \\
$B_{\rm K}$  	&          & 1.62 & 1.24 & 1.40 & 1.64 & 0.44 \\
$\tau_J$ 	& Myr      & 4.0 & $-$13.2 & 17.5  & 7.5 & 15.6 \\
$\tau_K$	& Myr      & 11.8 & 13.3 & 27.6 & 14.3 & 26.0 \\
$c_{\rm J}^*$ 	& pc/kpc   & 155. & 266. & 136. & 173. & 95. \\
$c_{\rm K}^*$ 	& pc/kpc   & 151. & 262. & 413. & 164. & 328. \\
$R_{\rm C}$ 	& kpc      & 6.34 & 6.95 & 5.96 & 6.85 & 6.44 \\
                     & & & & & &  \\ 
\tableline
\multicolumn{1}{l}{Dust:}                & & & & & & \\ 
$\kappa_{\rm V}$ & (MJy/sr)$^{-1}$  & 0.0162 & 0.0178 & 0.0116 & 0.0180  
								& 0.0188 \\
$k_{+}$ 	&     	& 3.95 & 4.17 & 4.06 & 4.02 & 3.38 \\
$k_{-}$ 	&     	  & 1.28 & 1.35 & 1.31 & 1.30 & 1.09 \\
$k_{\rm a}$  	&    	& 2.17 & 2.08 & 2.06 & {\bf 1.00} & 1.59 \\
$\rho_{\rm s} $ & MJy/sr/kpc 	 & 122. & 115. & 118. & 119. & 142. \\
$\rho_{\rm a}$  & MJy/sr/kpc 	 & 75. & 78. & 79. & 162. & 102. \\
                    & & & & & & \\
\tableline
\multicolumn{1}{l}{Miscellaneous: }      & & & & & & \\ 
$Z_\sun$ 	& pc          & 14.8 & 13.2 & 14.8  & 14.7 & 14.5 \\
$a_w$ 		& pc/kpc$^2$   & 24.5 & 19.4 & 22.0 & 29.1 & 25.6 \\
$Q_{J}$ 	& MJy/sr   & $-$0.0731 & $-$0.0585 & {\bf 0.0000} & $-$0.0687 
								& $-$0.0621  \\
$Q_{K}$ 	& MJy/sr   & $-$0.0798 & $-$0.0648 & {\bf 0.0000} & $-$0.0740 
								& $-$0.0620 \\
\enddata

\end{deluxetable}

\clearpage
%
%
\begin{deluxetable}{llccccc}
\tabletypesize{\footnotesize}
\tablewidth{0pc}
\tablecaption{Alternative Spiral models. \label{spmods}}
\tablehead{
\colhead{Parameter} & \colhead{units} &
\colhead{axisymmetric}   & \colhead{m=2 log}   &
\colhead{m=4 log} & \colhead{sheared} 
}
\startdata
Disk:         & & & & & &   \\ 
$\eta^0_{\rm J}$ & MJy/sr/kpc 	& 14.9 & 15.1 & 11.8 & 14.7  \\
$\eta^0_{\rm K}$ & MJy/sr/kpc 	& 12.6 & 12.5 & 9.9  & 11.6  \\
$r_*$ 	         & kpc		& 2.257  & 2.255  & 2.386  & 2.264    \\
$h_*$ 	         & pc		& 266.8  & 274.1  & 286.0  & 282.2  \\
$r_c$            & kpc		& 10.70  & 10.90  & 10.03  & 10.52   \\
              &  & & & & &   \\ 
\tableline
Spiral arms:    & & & & &   \\
sheared:    & & & & &   \\
$B_{\rm J}$    &              & -- & -- & -- & 0.86  \\
$B_{\rm K}$    &              & -- & -- & -- & 1.28  \\
$\tau_J$       &  Myr         & -- & -- & -- & 5.5   \\
$\tau_K$       &  Myr         & -- & -- & -- & 17.6    \\
$c_{\rm J}^*$  & pc/kpc       & -- & -- & -- & 142.  \\
$c_{\rm K}^*$  & pc/kpc       & -- & -- & -- & 143.  \\
$R_{\rm C}$    & kpc          & -- & -- & -- & 6.66   \\
logarithmic:  &  & & & & &   \\
$B$            &              & -- & 1.38    & 0.42    & --   \\
$p$            &  deg         & -- & 25.4    & 15.8    & --   \\
$R_0$          &  kpc         & -- & 5.38    & 2.56    & --    \\
$c_{\rm a}^*$  & pc/kpc       & -- & 138.    & 99.     & --  \\
                        &	&		& & & &    \\
\tableline
Dust:             	&   &   &  & & &  \\
$k_{+}$        &  		& 2.65   & 3.61   & 3.53   & 3.98   \\
$k_{-}$        &  		& 0.86   & 1.17   & 1.14   & 1.29   \\
$k_{\rm a}$    &    		& 1.00   & 2.19   & 11.25  & 2.64   \\
$\rho_{\rm s}$ & MJy/sr/kpc  	& 181.   & 133.   & 136.   & 121.   \\
$\rho_{\rm a}$ & MJy/sr/kpc   	& 162.   & 74.    & 14.    & 61.    \\
                & & & & & &  \\ 
\tableline
Miscellaneous:   & &  & & & &   \\
$\kappa_{\rm V}$ & (MJy/sr)$^{-1}$ & 0.0143  & 0.0172  &  0.0158  & 0.0180  \\
$Q_{J}$          &  MJy/sr         & -0.0533 & -0.0633 & -0.0640  & -0.0684 \\
$Q_{K}$          &  MJy/sr         & -0.0744 & -0.0760 & -0.0814  & -0.0744 \\
$Z_\sun$         &  pc       	   &  14.7   & 14.8    &  14.9    & 14.6    \\
$a_w$            & ${\rm pc/kpc}^2$ &  25.3   & 24.0    & 28.4     & 27.4 \\
$\chi^2$ for $|b|<3\deg$ &         & 65502   & 51547   & 53047    & 51452  \\
\enddata
\end{deluxetable}


%
%

\clearpage 

\plotone{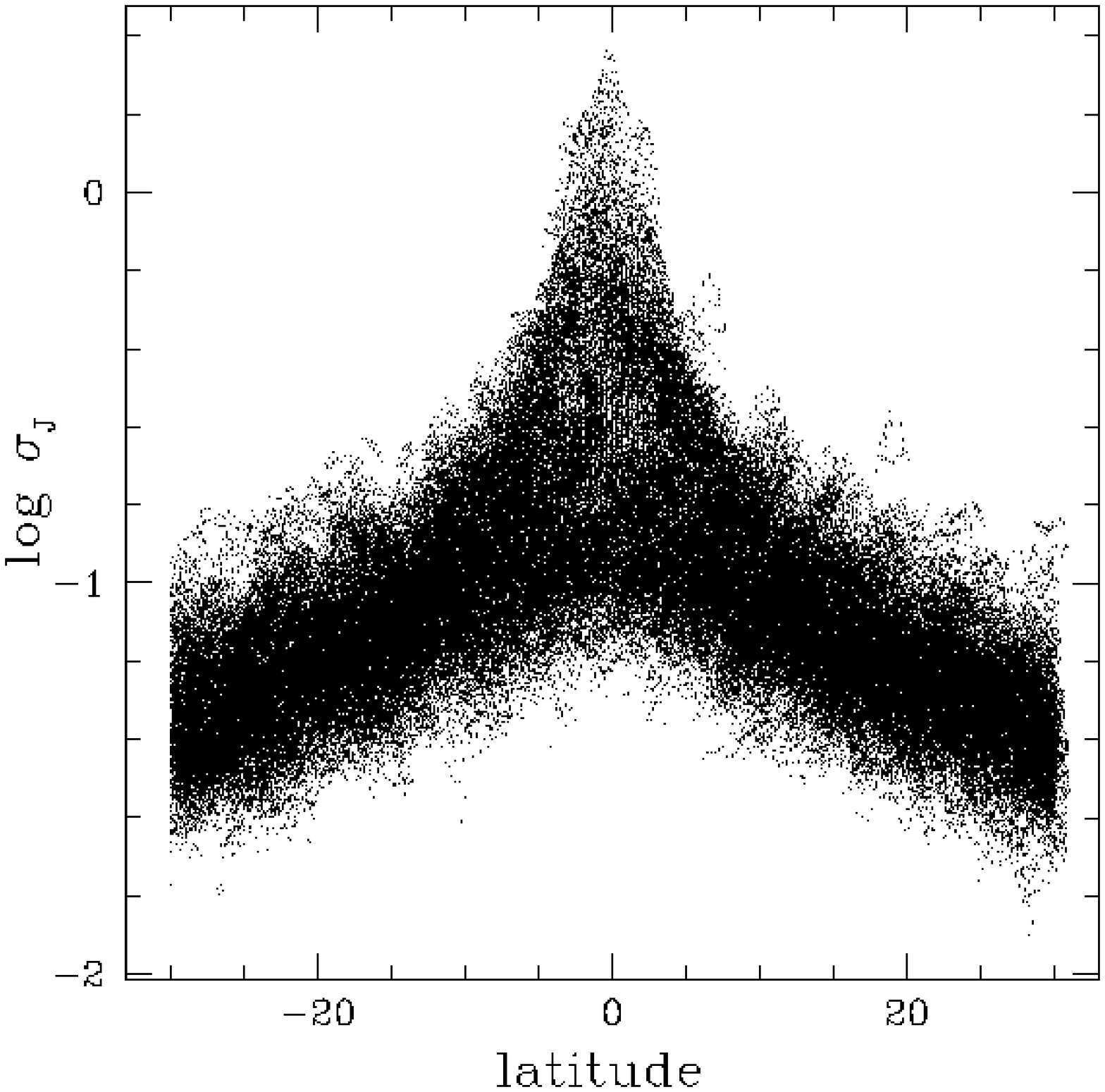}
\plotone{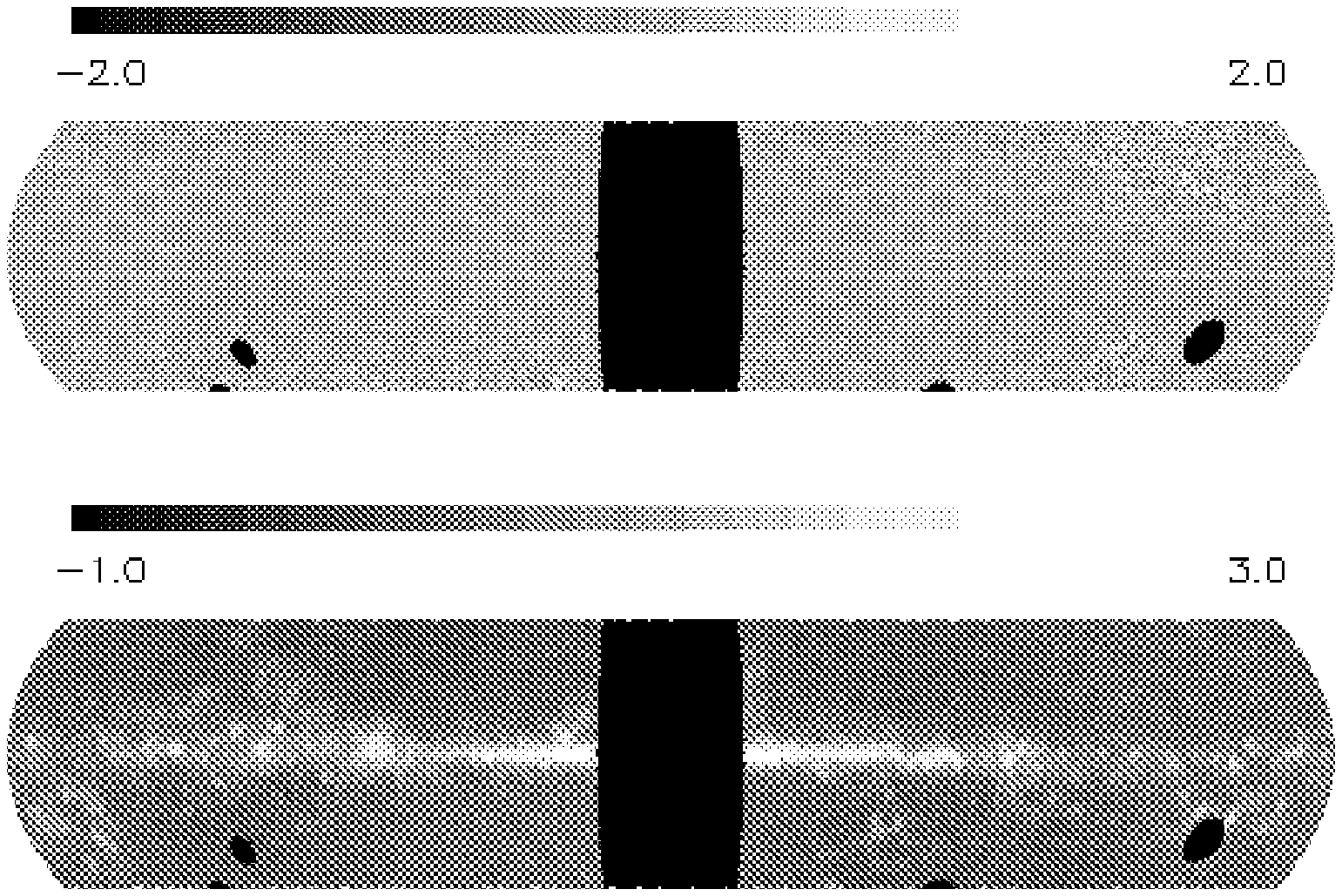}
\plotone{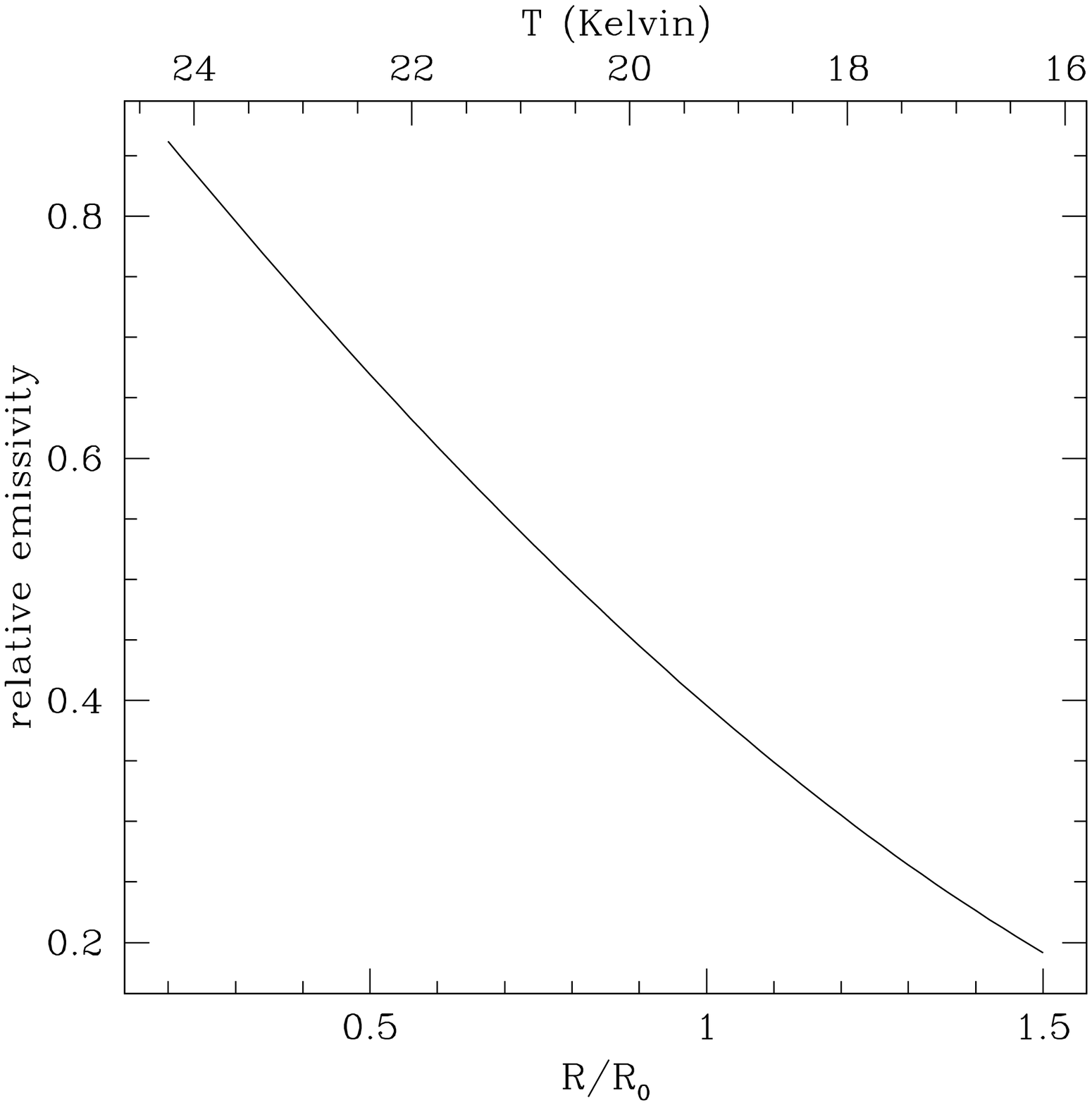}
\plotone{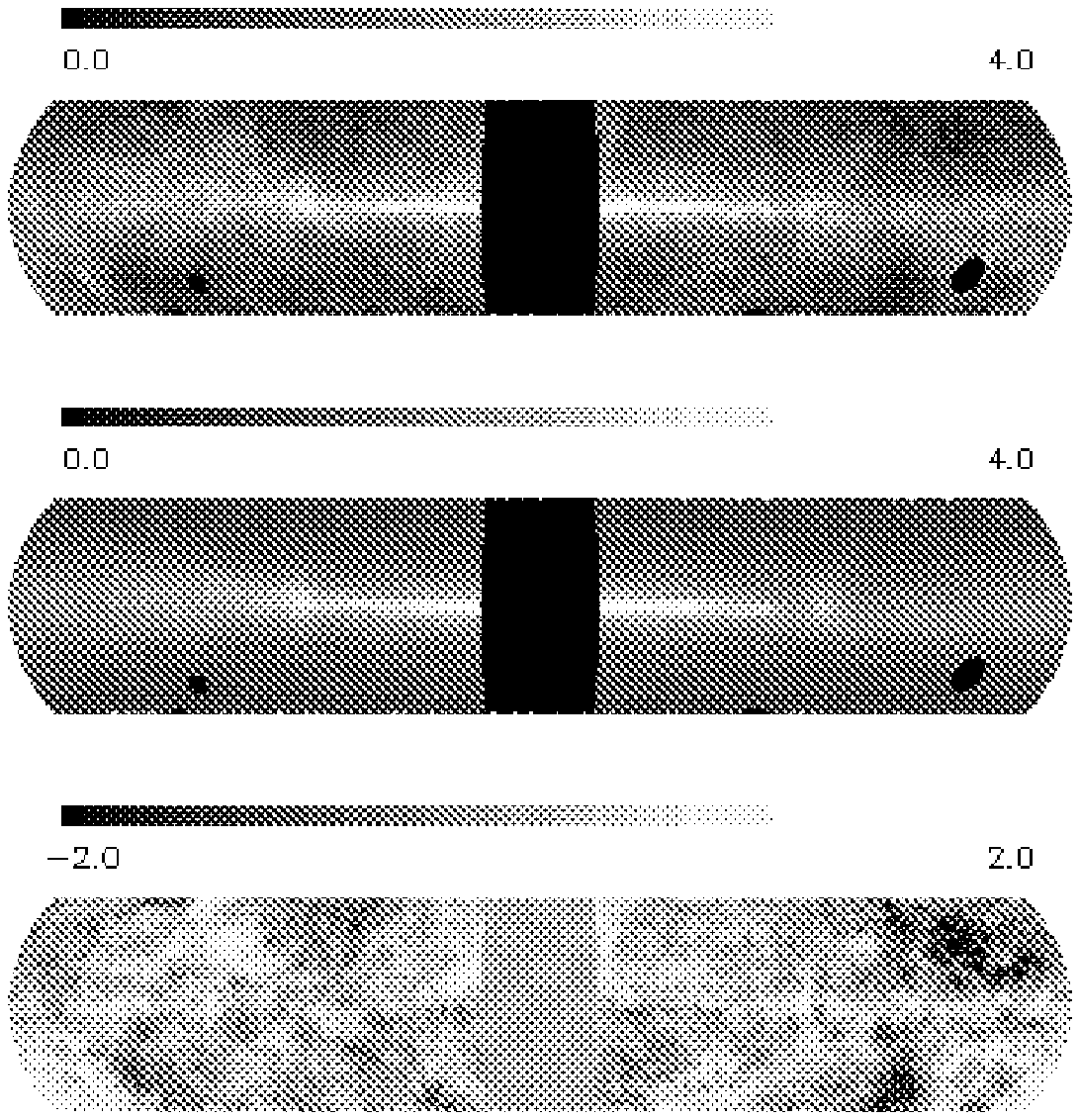}
\plotone{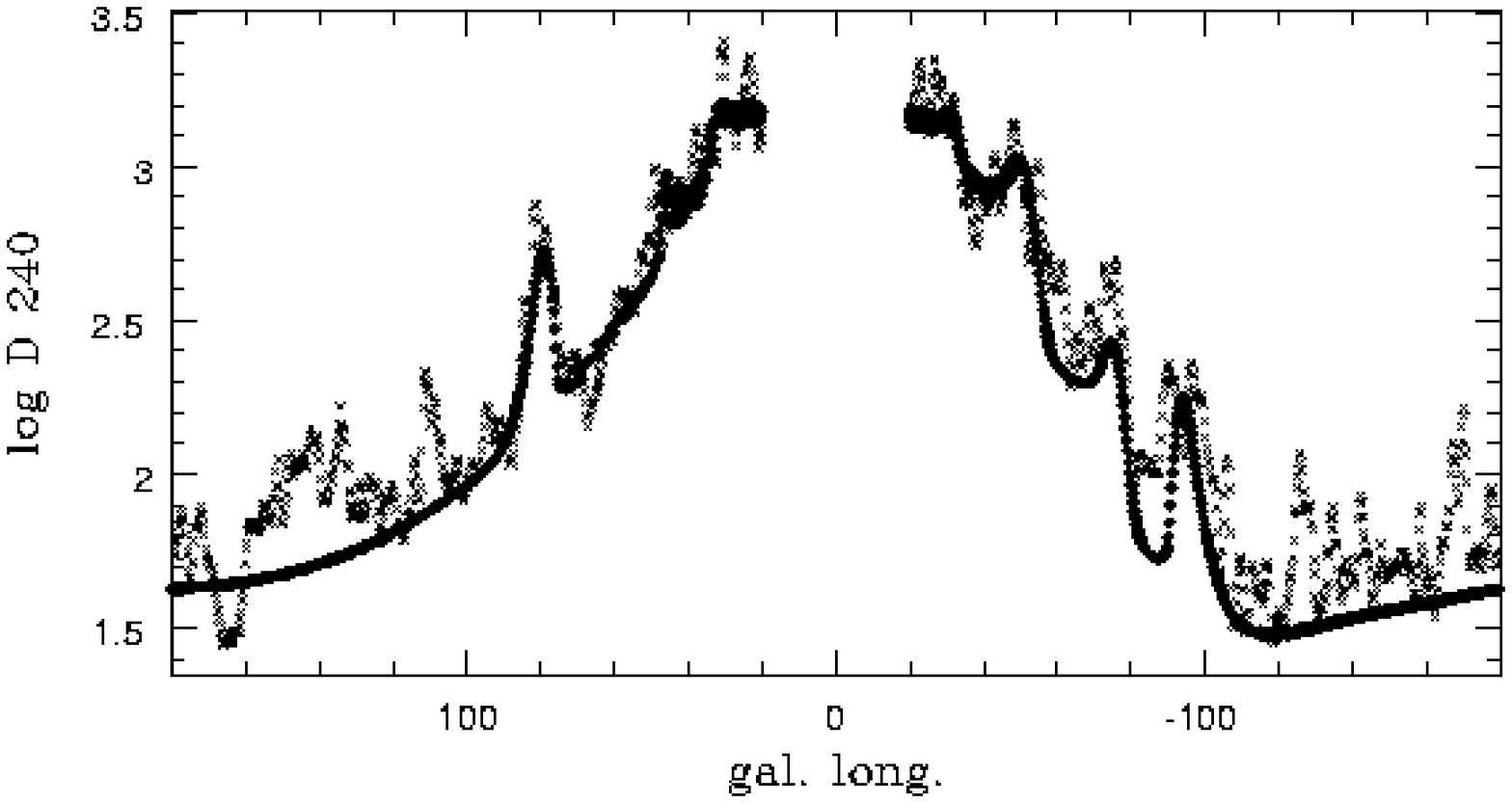}
\plotone{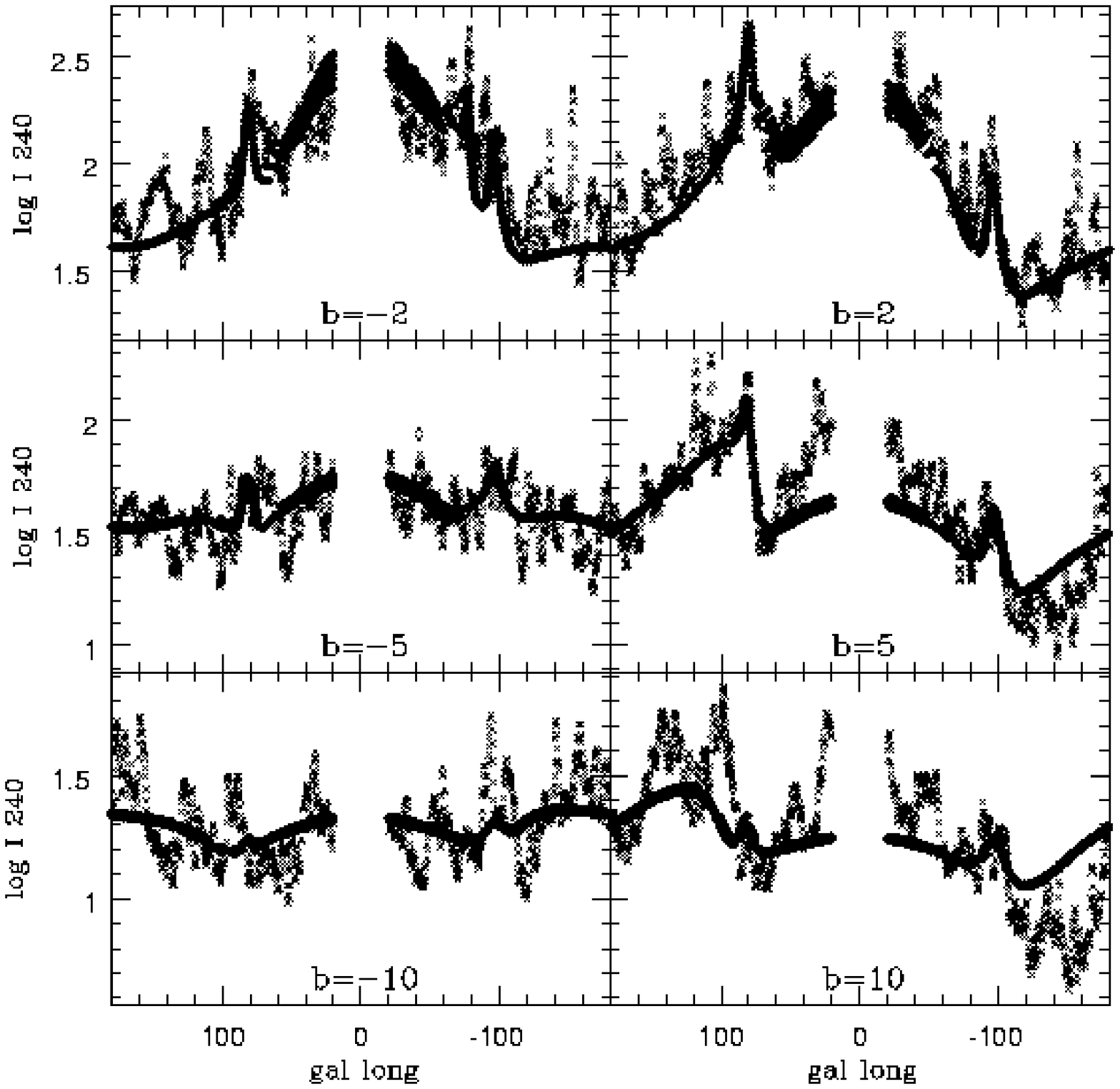}
\plotone{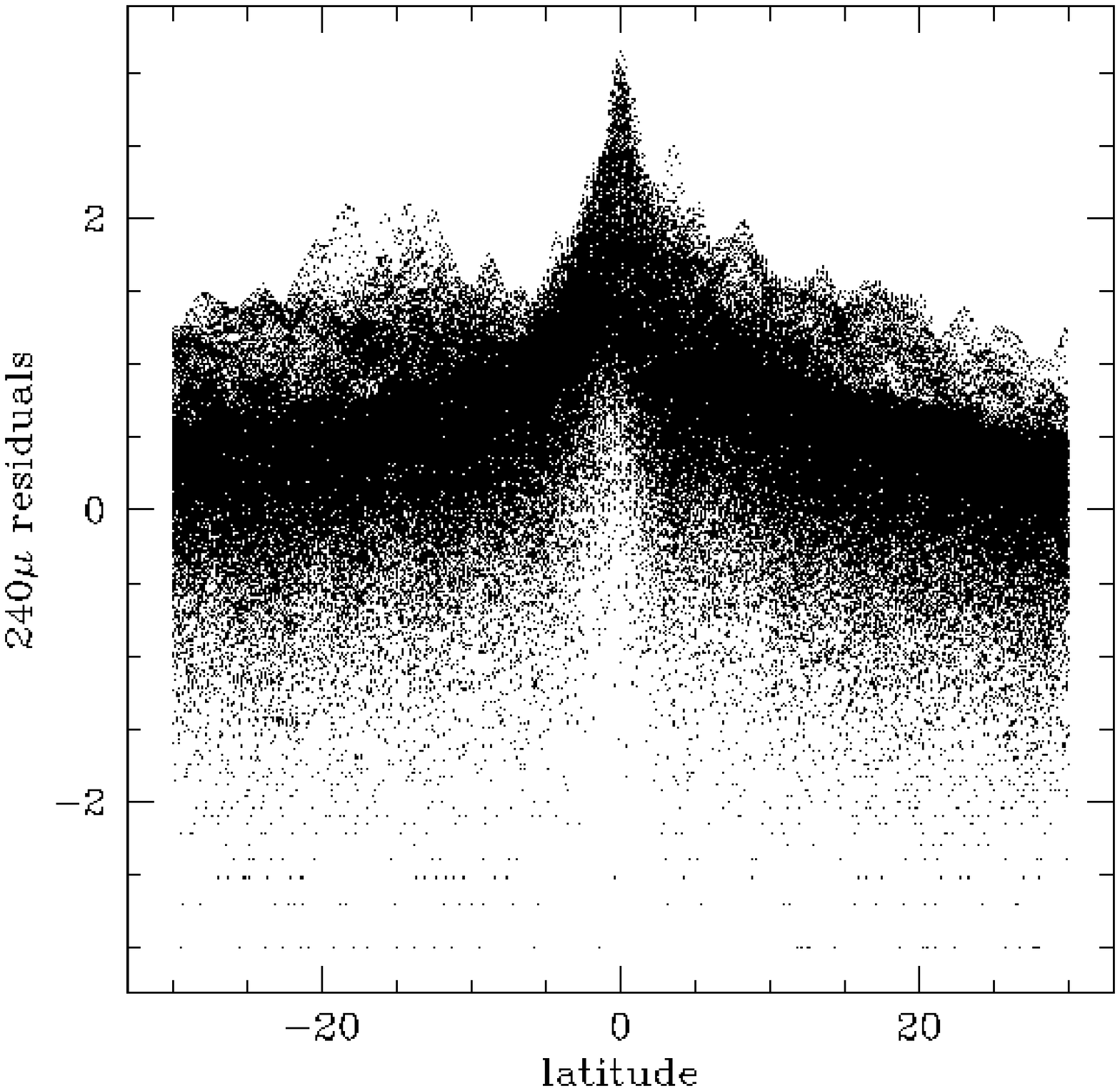}
\plotone{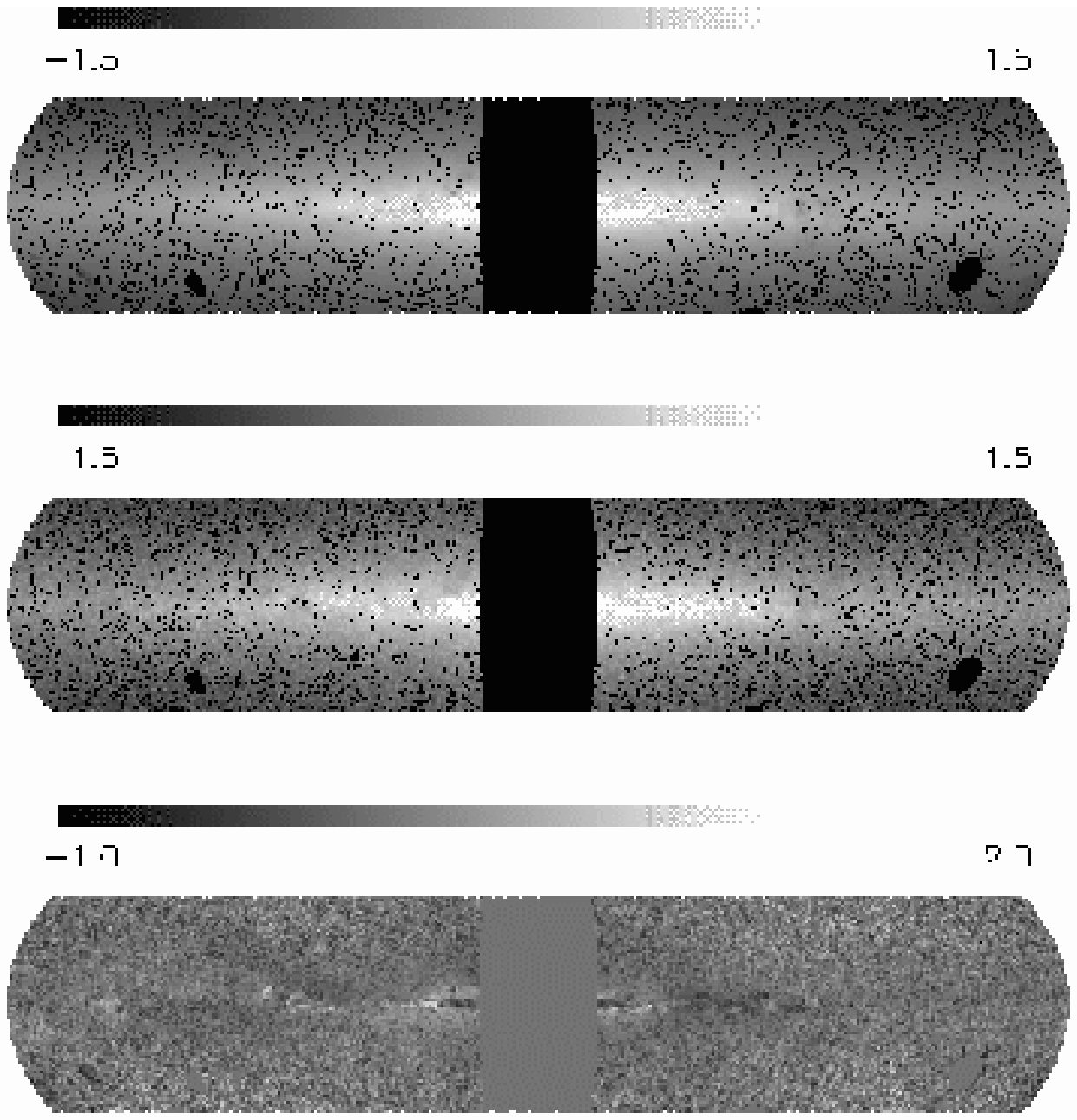}
\plotone{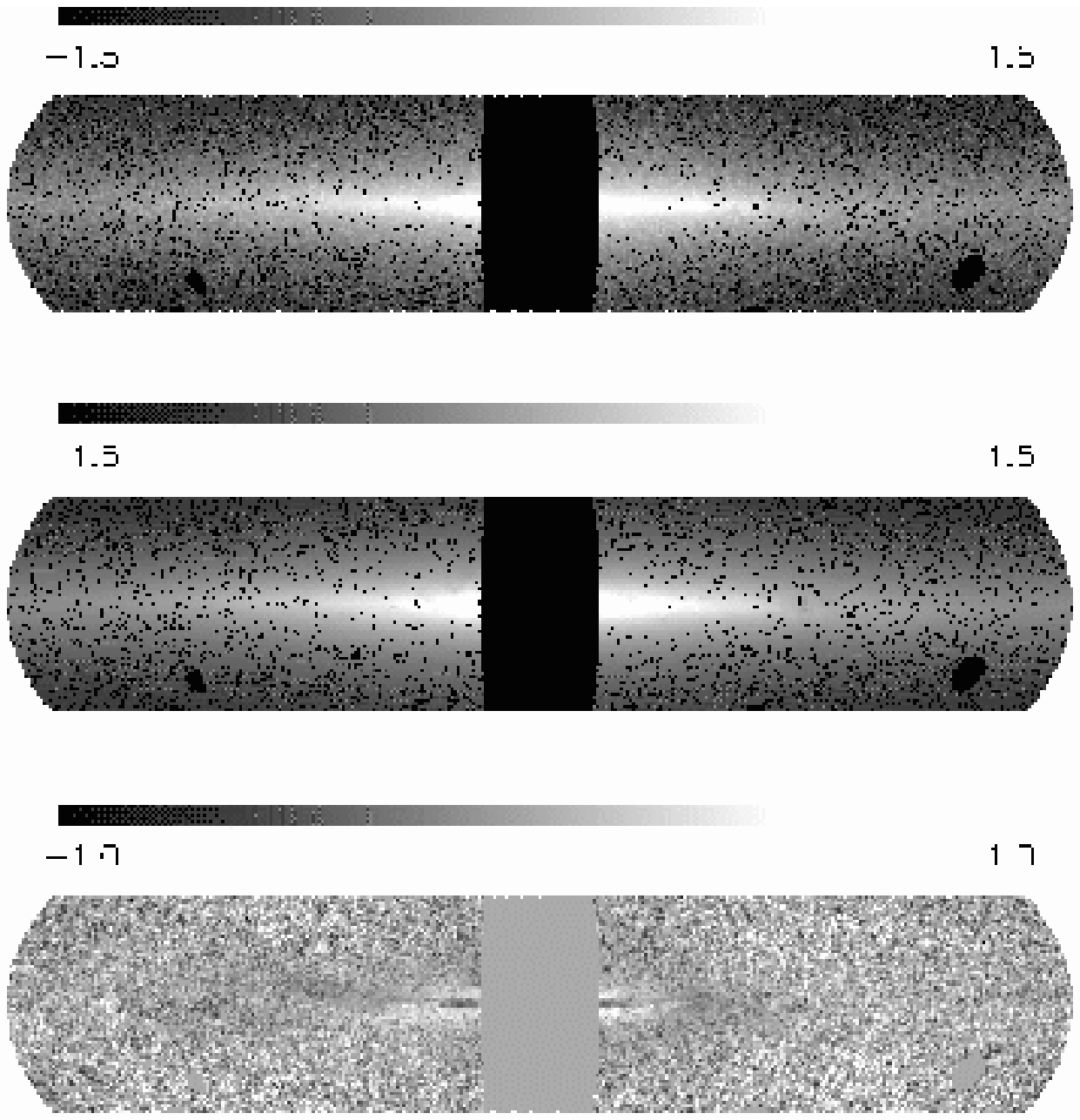}
\plotone{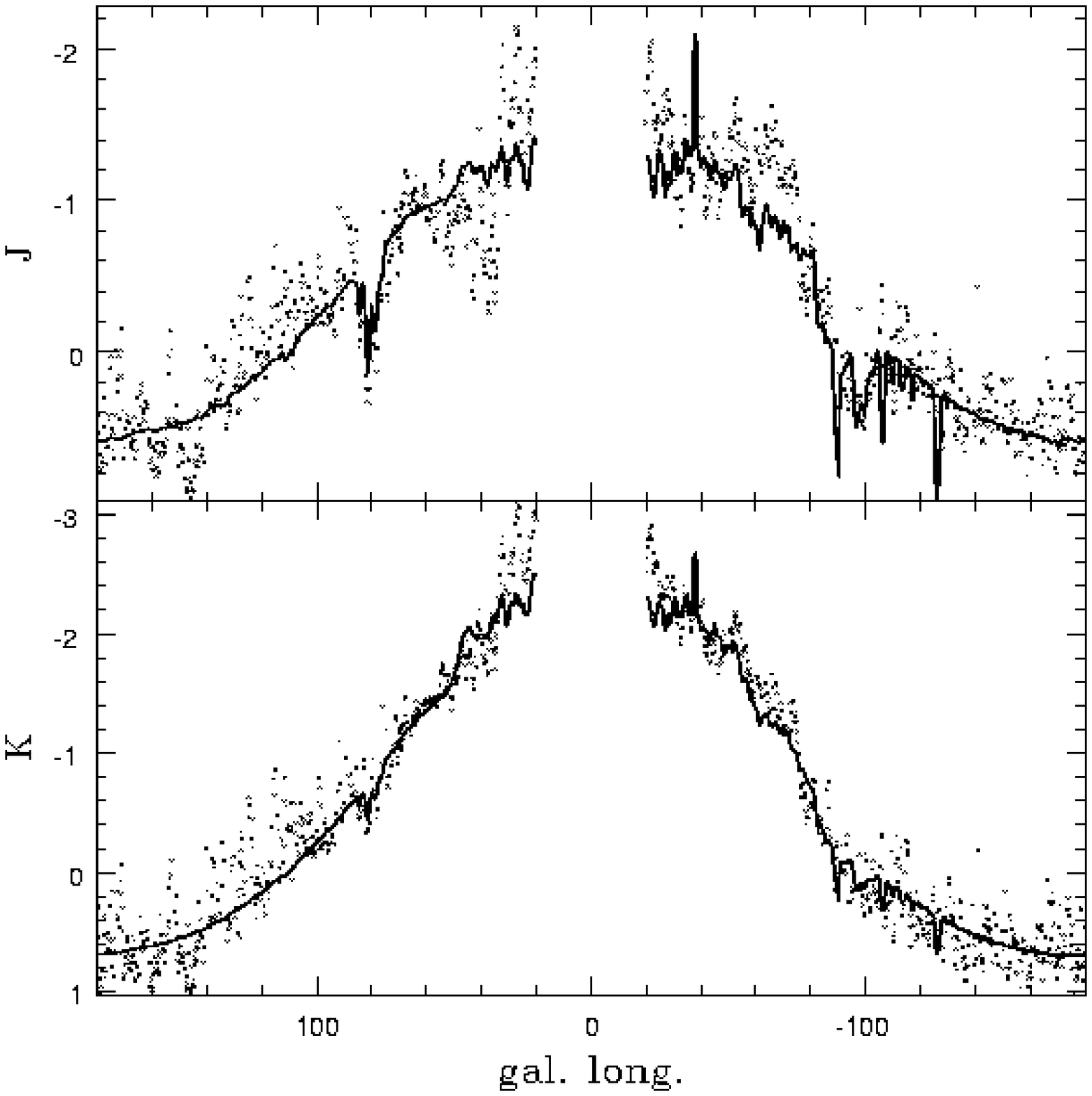}
\plotone{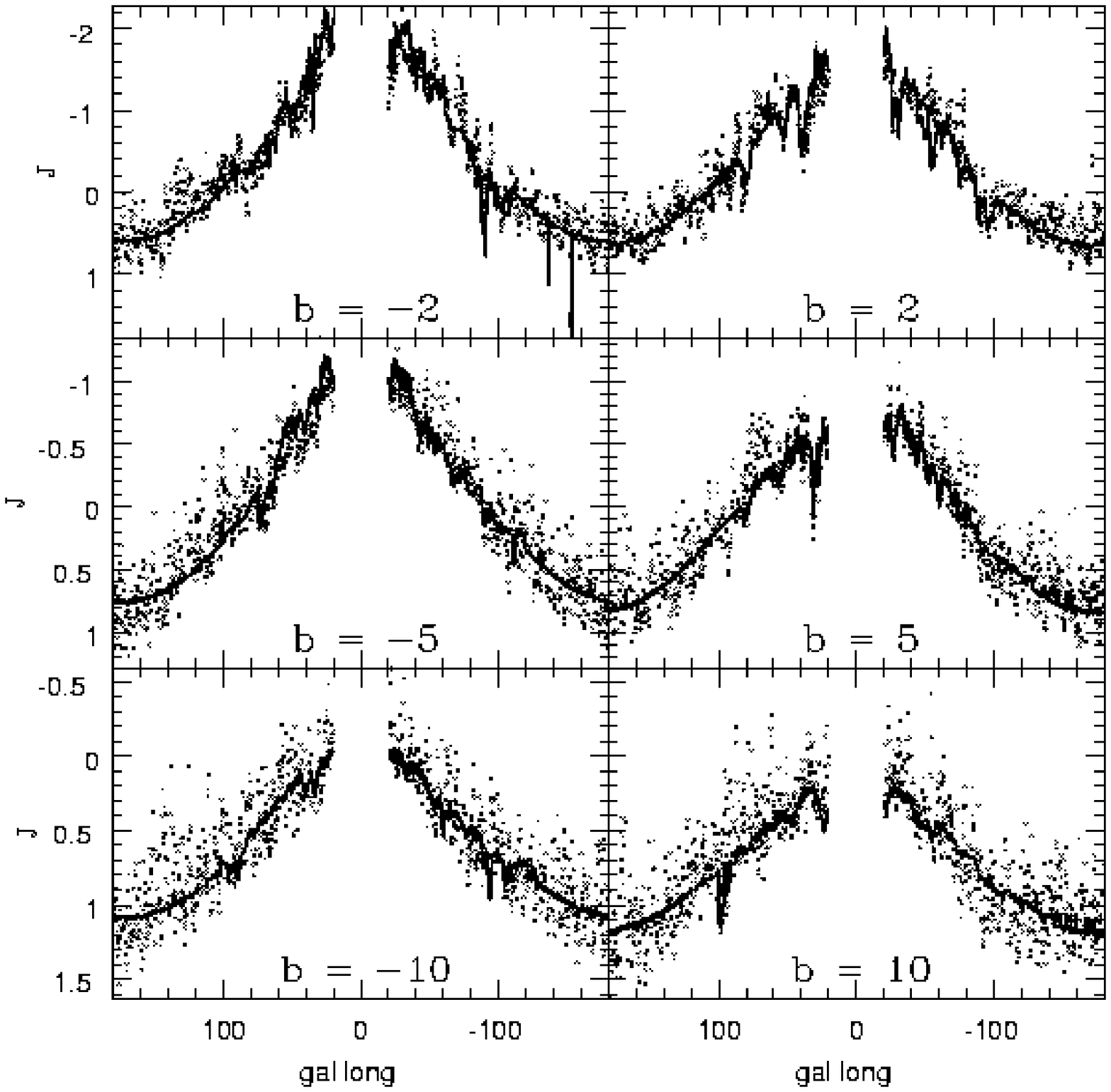}
\plotone{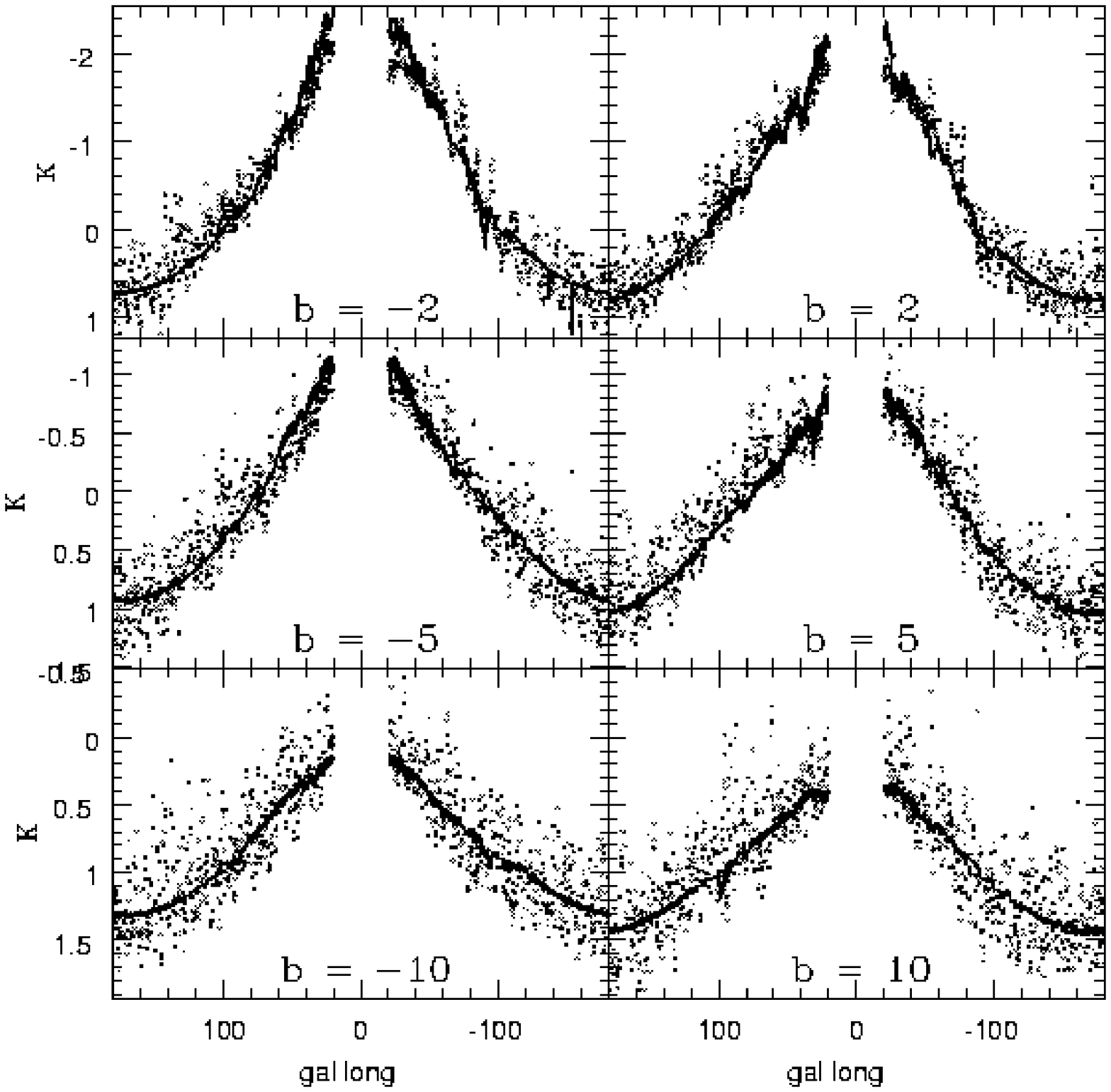}
\plotone{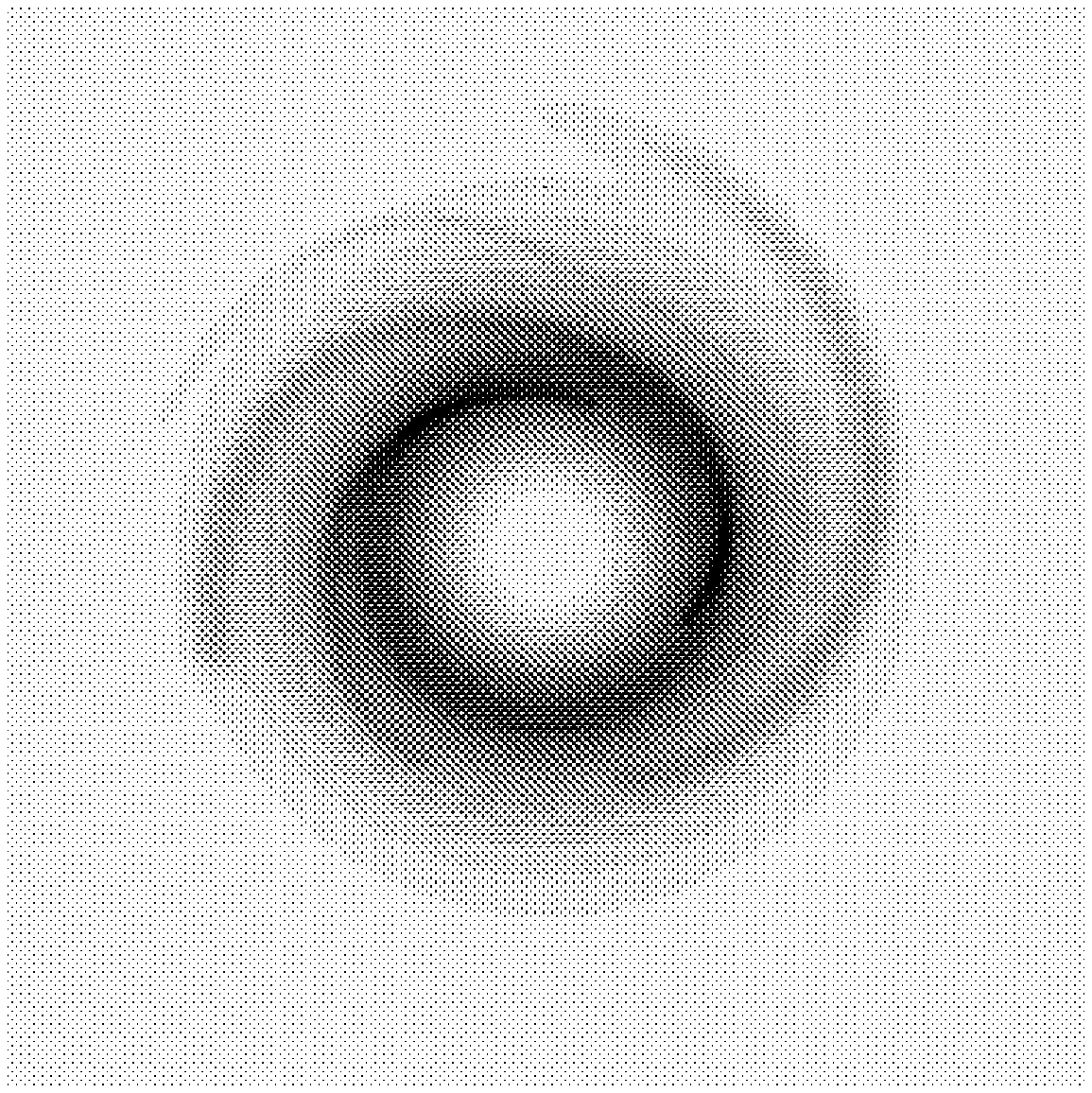}
\plotone{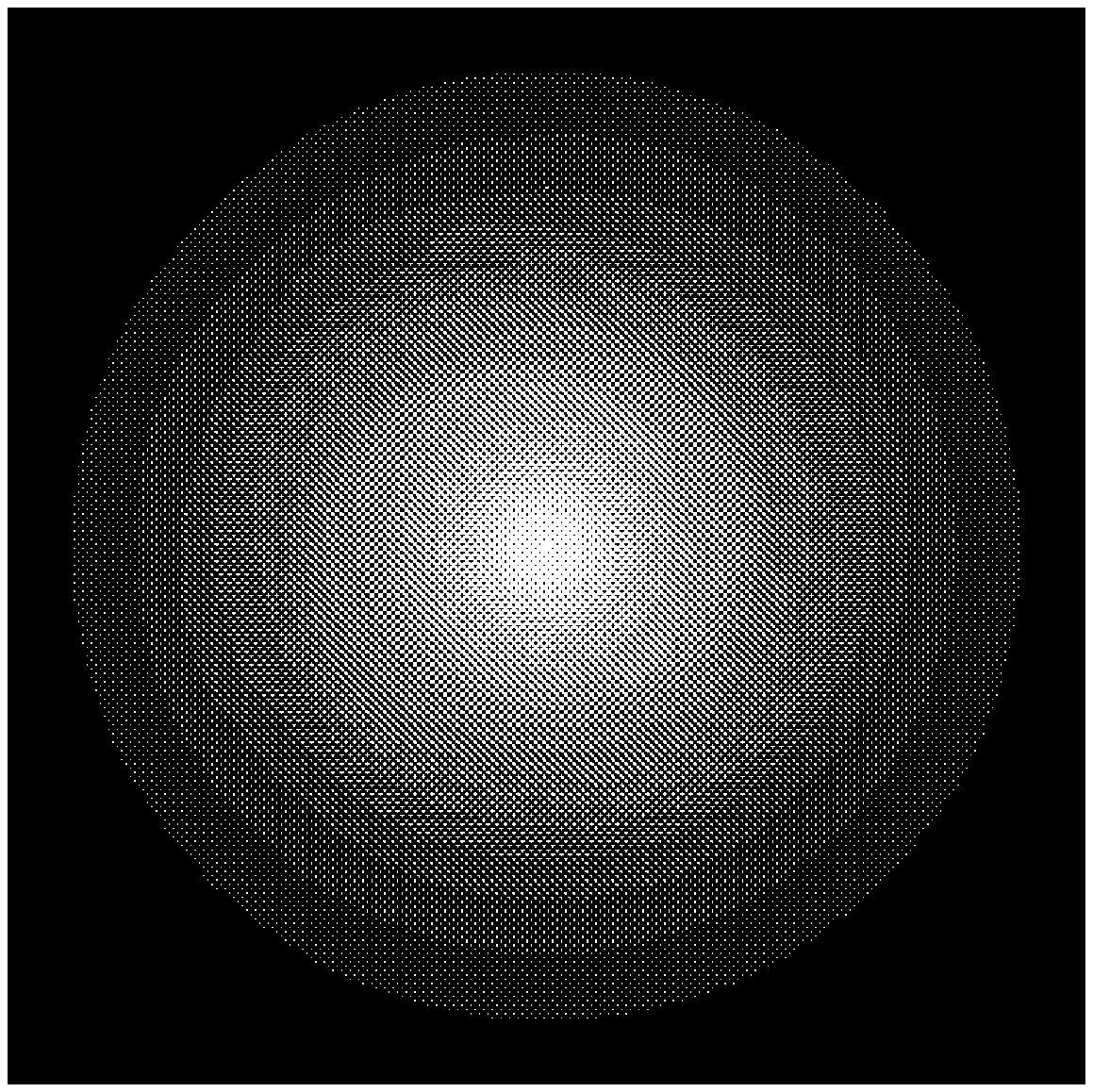}
\plotone{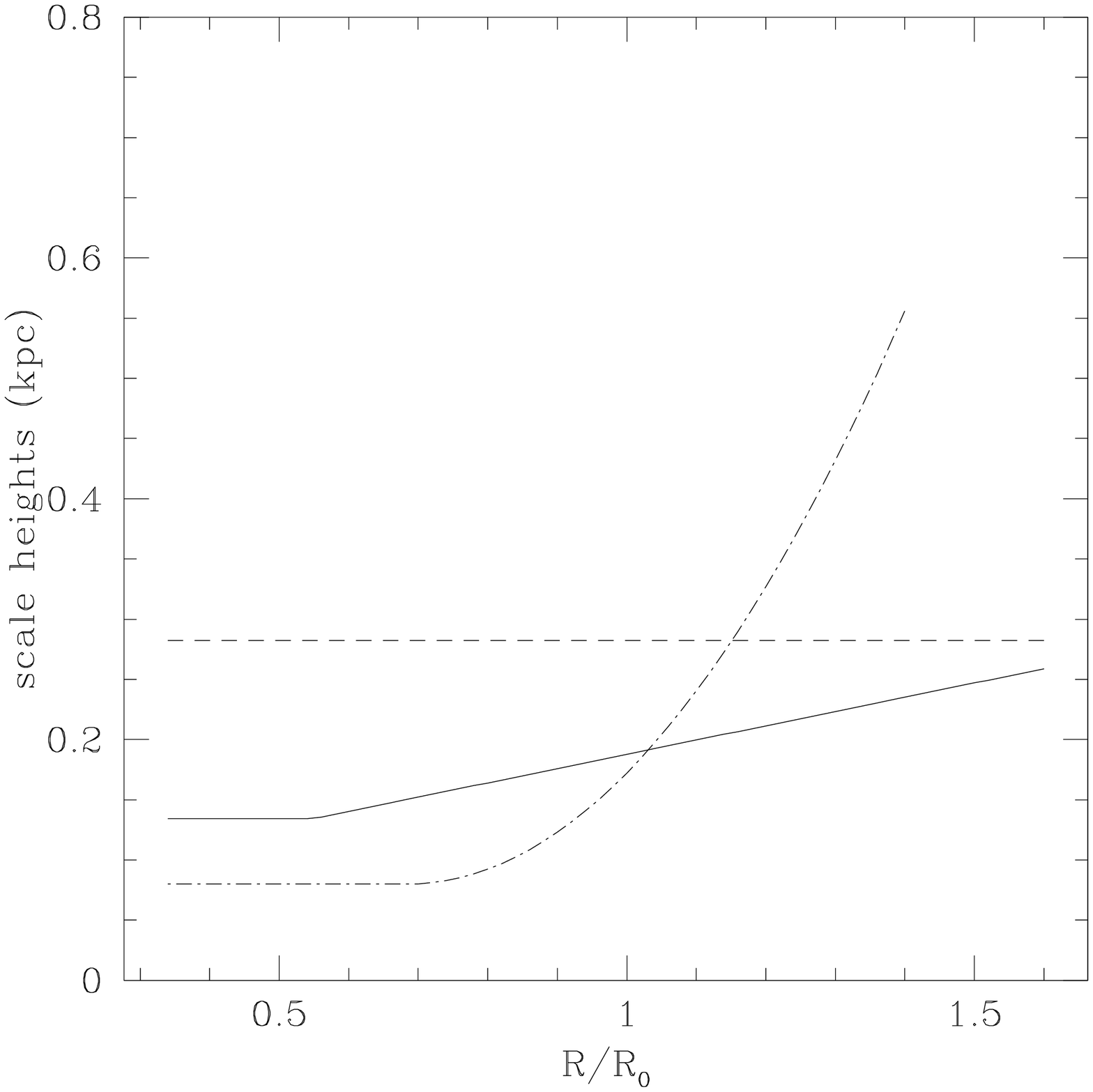}
\plotone{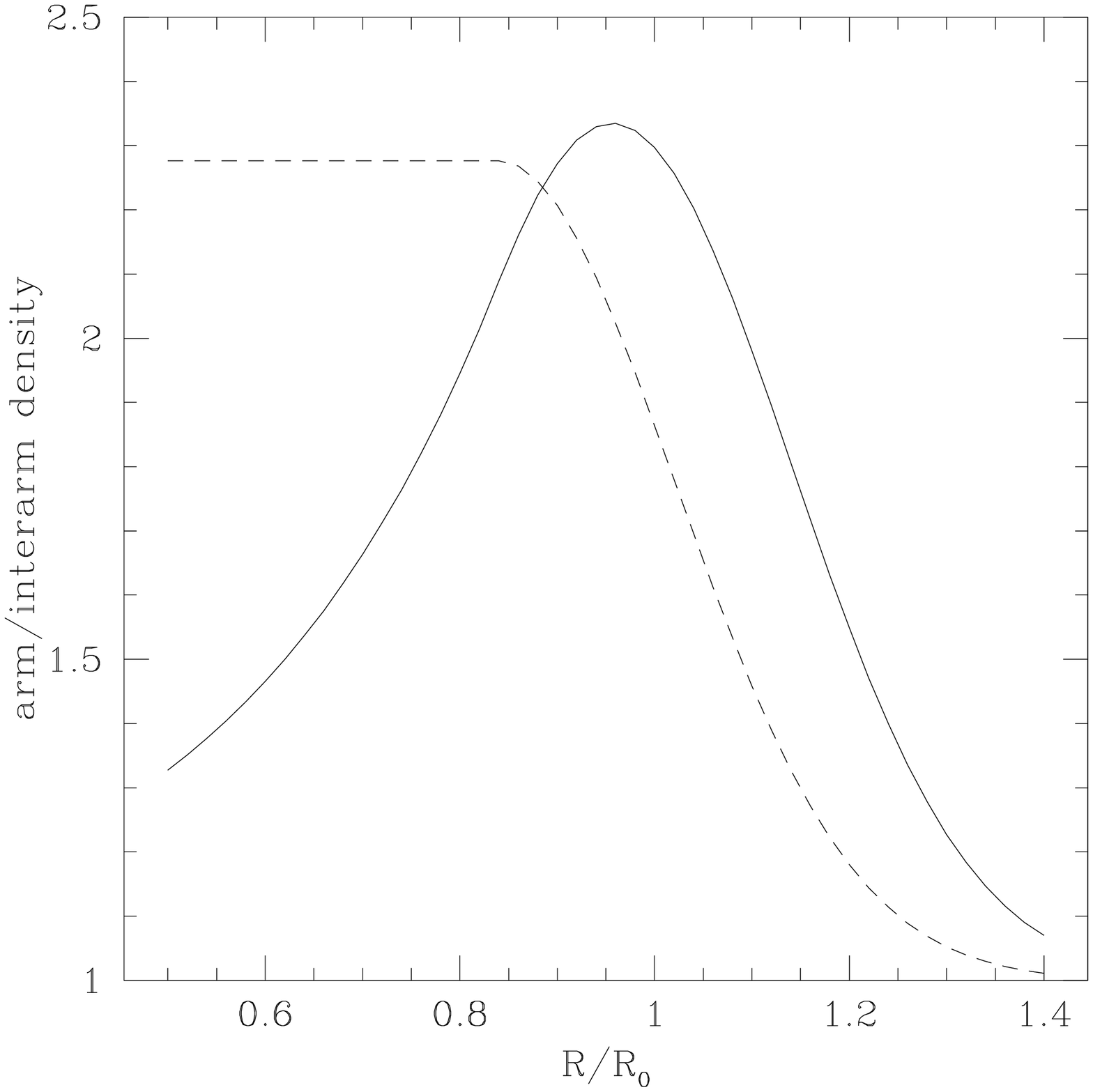}
\plotone{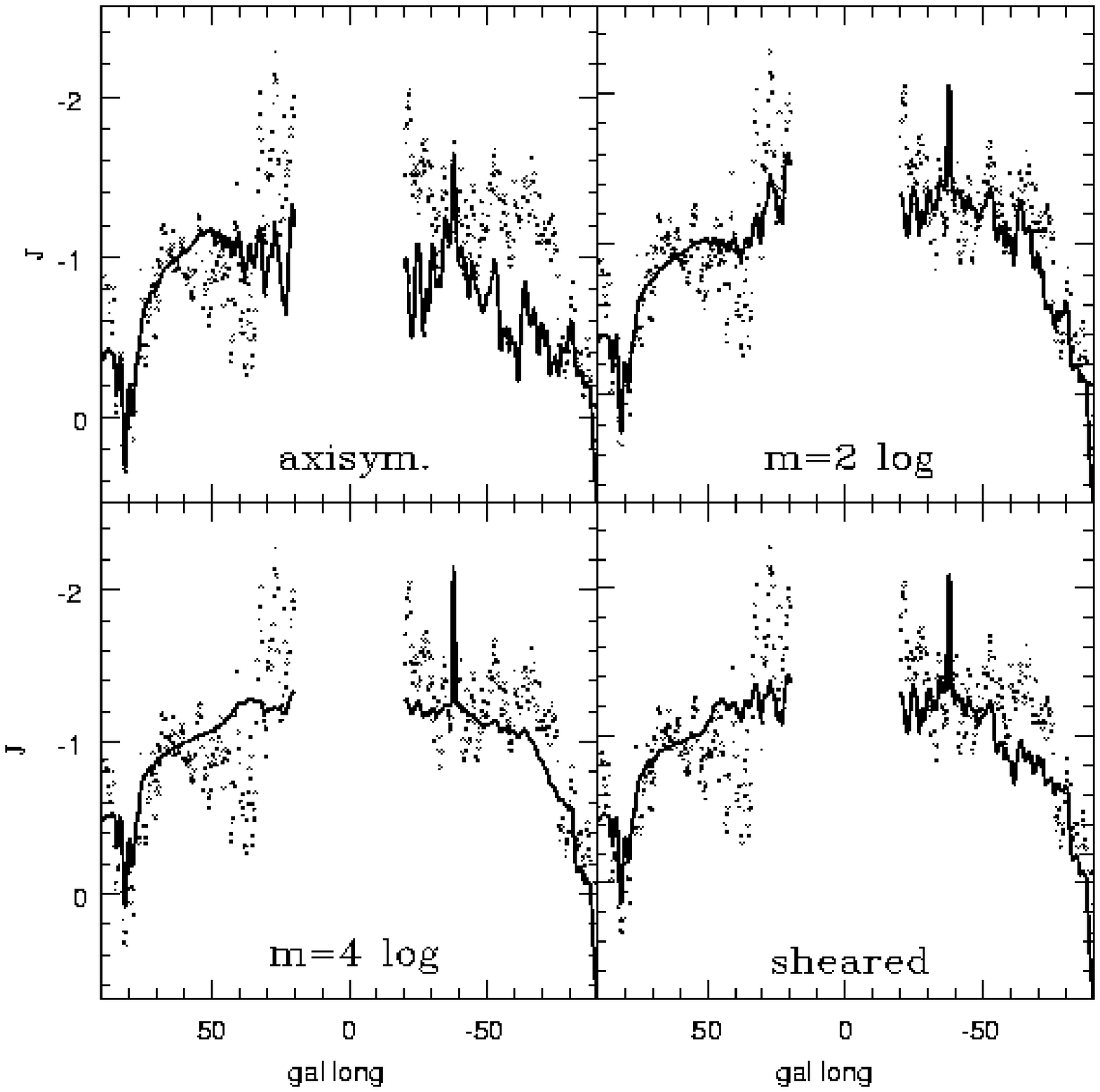}
\plotone{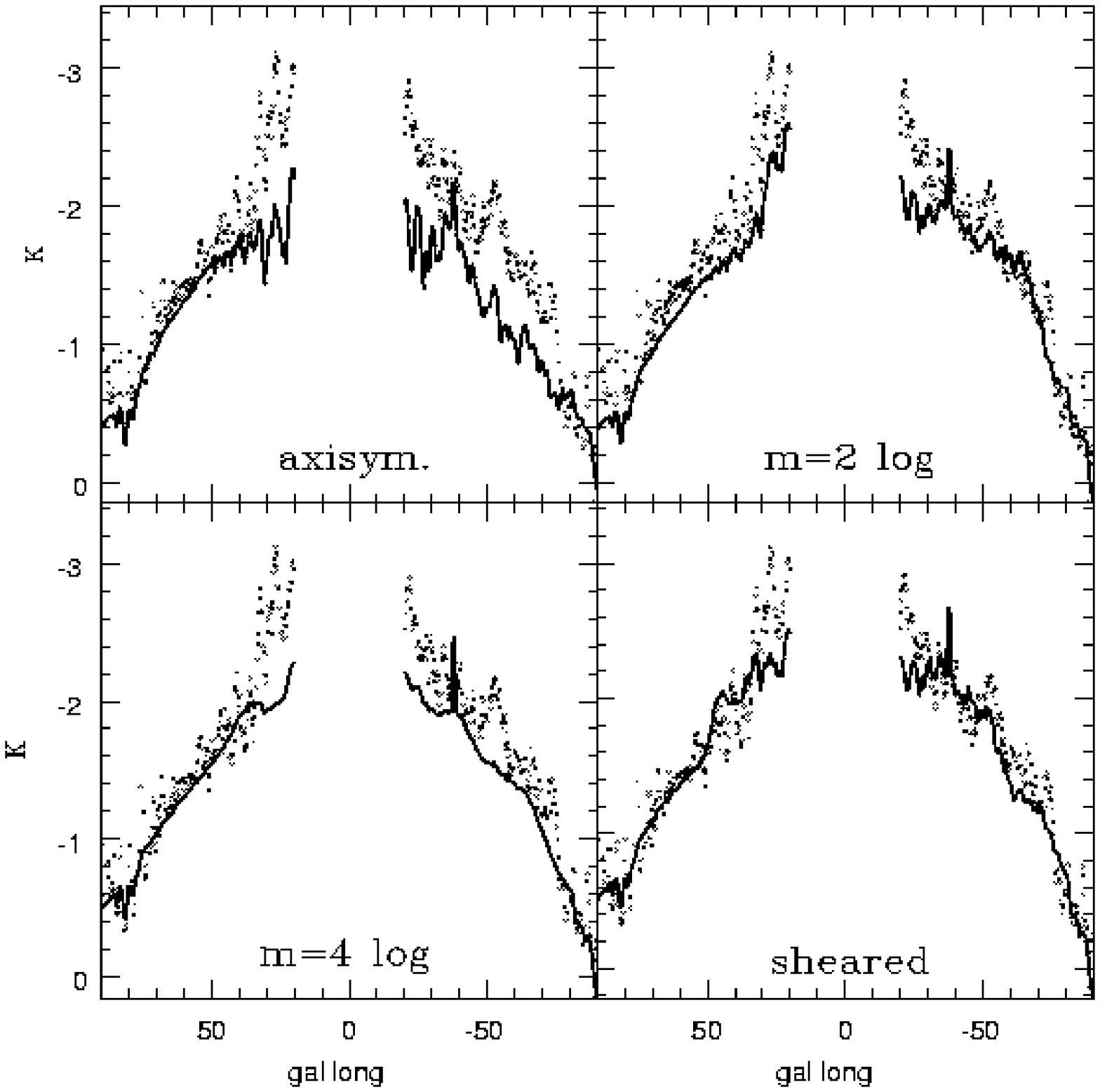}
\plotone{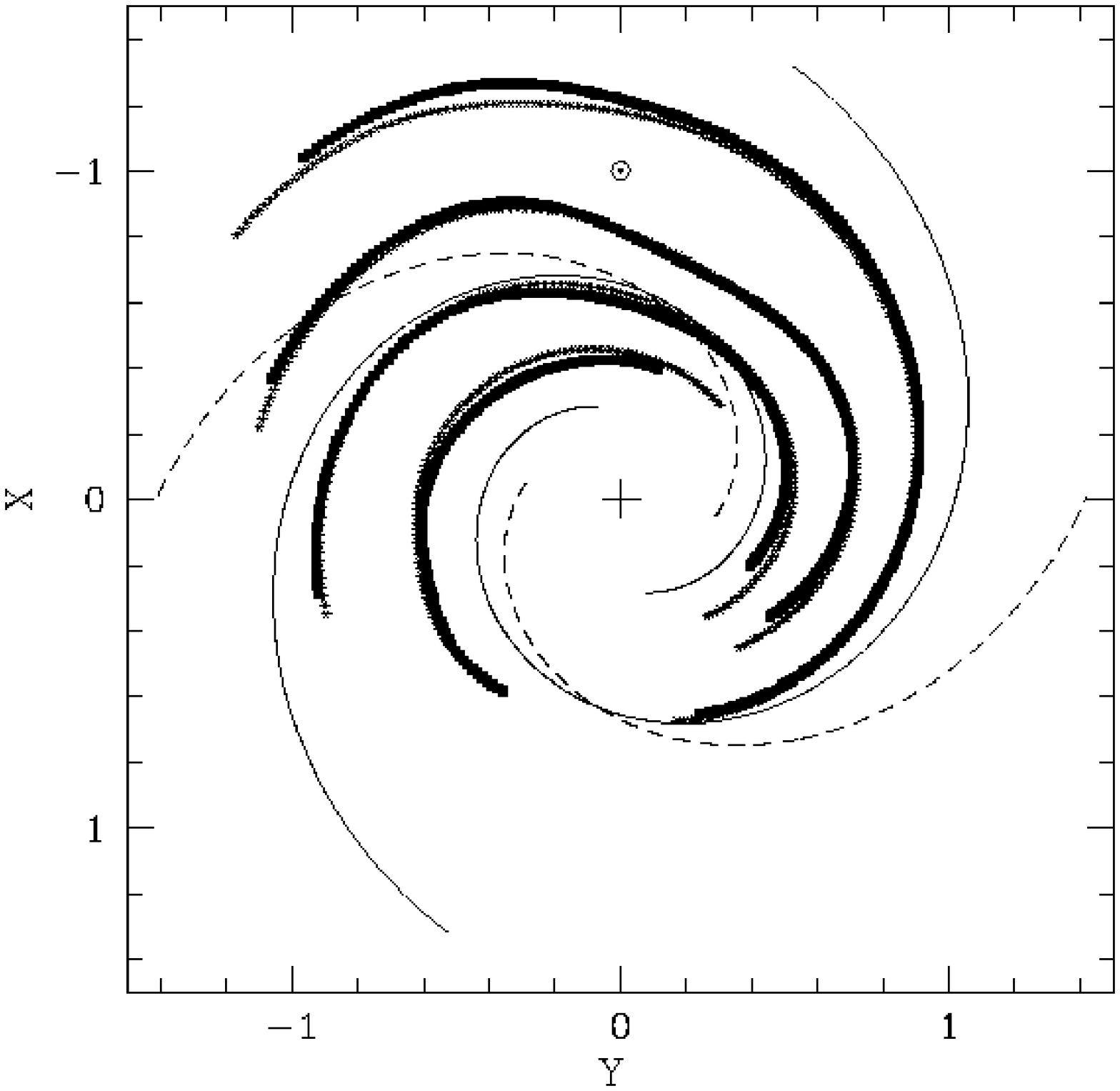}

\end{document}